\documentclass[reprint,aip,pop,graphicx,numerical,amsfonts,amsmath,amssymb,twocolumn]{revtex4-1}

\usepackage{bm}
\usepackage[dvips]{graphicx}
\usepackage{lineno}
\usepackage{color}
\usepackage{soul}
\usepackage{booktabs}
\usepackage{multirow}
\sethlcolor{yellow}

\newcommand{\biota}{\iota
                     \hskip-.15ex{\hbox to 0pt{\hss {\leavevmode
                     \hbox{\raise -.60ex \hbox{{\tt \'{}}}}}}}
                     \hskip.37ex{\hbox to 0pt{\hss {\leavevmode
                     \hbox{\raise -.50ex \hbox{{\tt \'{}}}}}}}}

\begin{document}

\preprint{00}

\title{
Improved linearized model collision operator for the highly collisional regime
}



\author{H. Sugama}
\affiliation{
National Institute for Fusion Science, 
Toki 509-5292, Japan
}
\affiliation{
Department of Fusion Science, SOKENDAI (The Graduate University for Advanced Studies), 
Toki 509-5292, Japan 
}

\author{S. Matsuoka}
\affiliation{
National Institute for Fusion Science, 
Toki 509-5292, Japan
}
\affiliation{
Department of Fusion Science, SOKENDAI (The Graduate University for Advanced Studies), 
Toki 509-5292, Japan 
}

\author{S. Satake}
\affiliation{
National Institute for Fusion Science, 
Toki 509-5292, Japan
}
\affiliation{
Department of Fusion Science, SOKENDAI (The Graduate University for Advanced Studies), 
Toki 509-5292, Japan 
}

\author{M. Nunami}
\affiliation{
National Institute for Fusion Science, 
Toki 509-5292, Japan
}
\affiliation{
Department of Fusion Science, SOKENDAI (The Graduate University for Advanced Studies), 
Toki 509-5292, Japan 
}

\author{T.-H. Watanabe}
\affiliation{
Department of Physics,
Nagoya University,  
Nagoya 464-8602, Japan
}


\date{\today}

\begin{abstract}
The linearized model collision operator for multiple species plasmas 
given by 
H. Sugama, T.-H. Watanabe, and M. Nunami 
[Phys.\ Plasmas {\bf 16}, 112503 (2009)]
is improved to be properly applicable up to the highly collisional 
regime. 
The improved linearized model operator 
retains conservation laws of particles, momentum, and energy 
as well as it  
reproduces the same friction-flow relations 
as derived by the linearized Landau operator so that 
this model can be used to correctly evaluate neoclassical 
transport fluxes in all collisionality regimes. 
    The adjointness relations and Boltzmann's H-theorem
are exactly satisfied by the improved operator 
except in the case of collisions between unlike particle species
with unequal temperatures where 
these relations and H-theorem still holds approximately 
because there is a large difference between 
the masses of the two species 
with significantly different temperatures.  
   Even in the unequal-temperature case,  
the improved operator can also be modified so as to 
exactly satisfy the adjointness relations while it causes 
the values of the friction coefficients to deviate from 
those given by the Landau operator.
   In addition, 
for application to gyrokinetic simulations of turbulent transport, 
the improved operator is transformed into the gyrophase-averaged form 
with keeping the finite gyroradius effect. 
\end{abstract}


\maketitle 



\section{INTRODUCTION}

Coulomb collisions are the main mechanism which causes 
classical and neoclassical transport in 
magnetically confined 
plasmas.~\cite{RHH,Hinton,H&S,Balescu,Helander} 
Even though plasma confinement is generally dominated by 
turbulent transport rather than by collisional transport, 
collisions still have impacts on structures of 
phase-space distribution functions of particles,  
growth rates of instabilities, 
and micro/macroscopic profiles of plasma flows 
so that they indirectly influence 
turbulent transport processes 
as well.~\cite{Horton,Idomura,W&S2004,H&R1999,Lin_PRL,Nakata_PRL} 
Also, transport processes of 
heavy impurities with high charge numbers which penetrate 
from the edge into the core region are greatly affected by 
Coulomb collisions.~\cite{Casson,Helander_PRL,Dux,Yamoto} 
So far, there have been numerous works on model collision 
operators~\cite{HScollision,Dimits,Lin1995,Wang,Catto,Xu,Abel,Sugama2009,Brizard2004,Madsen,Burby,Sugama2015,Esteve,Hirvijoki,Sugama2017}
for application to 
theoretical and numerical studies of plasma transport. 

A well-established Coulomb collision term is given by 
the Landau operator~\cite{HM}  
which is nonlinear for like-species collisions
or bilinear for unlike-species 
collisions. 
The linearized Landau operator~\cite{Belli2012,Landreman2012,Pan} 
obtained by perturbatively 
expanding the distribution 
functions about the local Maxwellian is more tractable than 
the full Landau operator~\cite{Takizuka,Nanbu,XGC} 
 and the former is preferred to 
be used for transport studies when the deviation from 
the Maxwellian is sufficiently 
small. 
Since the field particle part of the linearized Landau operator 
is not as easy to evaluate as its test particle part, 
several linearized model collision operators have been proposed, 
in which simplified versions of the field particle part are 
used.~\cite{Dimits,Lin1995,Wang,Catto,Xu,Abel,Sugama2009} 
As an example, Sugama {\it et al}.~\cite{Sugama2009} presented 
a linearized model collision operator for multiple ion species plasmas 
which conserves particles, momentum, and energy, 
and satisfies adjointness relations and 
Boltzmann's H-theorem even for collisions 
between unlike particle species 
with unequal temperatures. 
This model called the Sugama operator has been successfully applied 
to studies of neoclassical and 
turbulent transport in relatively low collisional 
regimes.~\cite{Nakata,Nunami,Satake,Idomura2016,Candy,Belli2017,Maeyama,GENE}  

The difference between the field particle part of the Sugama operator and 
that of the exact linearized Landau operator is anticipated to increase 
in a highly collisional regime. 
Even in very-high-temperature fusion plasmas 
like the ITER plasma,~\cite{Dux,Yamoto,ITER}  
minority impurity ions such as tungsten are considered to remain 
in the Pfirsch-Schl\"{u}ter regime even though bulk 
ions and electrons are in the banana regime. 
For such a case, it is necessary to use a collision model 
which is accurate in all collisionality regimes. 
In this work, the Sugama operator is improved to present 
the new linearized model collision operator, 
which is properly applicable to all cases from low to high collisionality. 
The improved model is constructed so as to 
give exactly the same friction-flow relations 
as those derived from the linearized Landau operator.
Therefore, it can be used in drift kinetic simulations  
to accurately evaluate neoclassical transport fluxes in all collisionality regimes. 
Then, it is noted that
the exact friction-flow relations no longer rigorously keep 
the symmetry property in the case of collisions between unlike particle species
with unequal temperatures, where 
neither the improved model operator 
nor the linearized Landau operator 
is completely self-adjoint.  
Since the self-adjointness is practically 
useful for analytical or numerical derivation 
of the Onsager symmetric neoclassical transport 
coefficients,~\cite{RHH,Hinton,H&S,Balescu,Helander,DKES,Taguchi,Sugama1996,Sugama-Nishimura} 
further modification of the improved model 
for the unequal-temperature case is considered in the present paper 
to restore the adjointness relations by relaxing 
the accuracy of the friction-flow relations. 
In addition, the improved collision operator in the form suitable for application 
to gyrokinetic simulations of turbulent transport is derived by 
taking the gyrophase average 
with the finite gyroradius effect taken into account.

It is instructive to note here that 
Hirshman and Sigmar~\cite{HScollision} presented a linearized model collision operator which is similar to ours in that spherical harmonic functions and Laguerre polynomials  are used to expand distribution functions as well as key properties of the original linearized Landau collision operator are retained. 
In their work,~\cite{HScollision} 
an elegant and skillful method of constructing novel basis functions 
is presented to approximate both test and field particle 
operators including spherical harmonic functions of all degree numbers ($l$'s)
although an explicit expression of their model collision operator is given in their paper only for the case where spherical harmonic functions of degrees $l > 2$ are dropped.
To satisfy conservation laws of momentum and energy, 
the field particle part of the Sugama collision operator contains 
the $l  = 0$ and $1$ parts which are expressed using the test particle part  
and take similar forms to those of the Hirshman-Sigmar operator.  
In the present paper, the improved Sugama collision operator is given by adding the correction terms into only the $l = 1$ spherical harmonic component of the original Sugama operator in order to correctly reproduce the friction-flow relations 
which determine collisional transport and influence turbulent transport 
through controlling micro/macroscopic plasma flow profiles. 
However, the procedures shown in the present work can be extended 
to give correction terms to all other spherical harmonic components. 
It is also shown by Abel {\it et al}.~\cite{Abel} that,  
when the Hirshman-Sigmar model operator is transformed to 
its gyrophase-averaged form for application to the gyrokinetic equation, 
the problematic gyroradius dependence appears 
in the energy diffusion term in the test particle operator. 
Therefore, for the gyrokinetic case, our model operator  
is more favorable than the Hirshman-Sigmar model operator. 

The rest of this paper is organized ad follows. 
In Sec.~II, 
we briefly explain the Landau collision operator and its linearization, 
from which the associated matrix elements are defined to obtain 
the friction coefficients entering the friction-flow relations. 
Then, after reviewing the definition and properties 
of the original Sugama operator in Sec.~III, 
its improved version is presented in Sec.~IV, where we write down 
the correction term to reproduce the same matrix elements and 
friction coefficients as given by the linearized Landau operator. 
In Sec.~V, 
the improved operator is expressed in the form suitable for 
gyrokinetic equations. 
Finally, conclusions are given in Sec.~VI. 
In Appendix~A, 
a collisional energy transfer rate between unlike species  
with unequal temperatures is estimated depending on the ratio between 
the masses of the two species. 
In Appendix~B, effects of unequal temperatures of colliding particle species 
on the adjointness relations and matrix elements associated with 
the linearized Landau operator are discussed. 
The detailed expressions of the matrix elements 
are shown in Appendix~C.
In addition, 
Appendix~D presents
a modified version of the improved operator which exactly  
satisfies the adjointness relations even for collisions between 
unlike particle species with unequal temperatures although 
it consequently makes the values of the friction coefficients 
deviate from those given by the Landau operator.

\section{LANDAU COLLISION OPERATOR AND FRICTION-FLOW RELATIONS}

The Landau operator 
for collisions between particle species $a$ and $b$ 
is written as~\cite{HM}
\begin{eqnarray}
\label{CLandau}
C_{ab}  (f_a, f_b) 
&\equiv & 
- \frac{2\pi e_a^2 e_b^2 \ln \Lambda}{m_a}
\frac{\partial}{\partial {\bf v}}
\cdot \left[
\int d^3 v' \;
{\bf U} ( {\bf v} - {\bf v}' )  
\right. \nonumber \\ & & 
\left. \mbox{} 
\cdot 
\left\{ 
\frac{f_a ({\bf v})}{
m_b
}
\frac{\partial f_b ({\bf v}')}{\partial {\bf v}'}
- \frac{f_b ({\bf v}')}{
m_a
}
\frac{\partial f_a ({\bf v})}{\partial {\bf v}}
\right\}
\right]
, 
\hspace*{5mm}
\end{eqnarray}
where 
\begin{equation}
{\bf U} ( {\bf v} - {\bf v}' ) \equiv 
\frac{| {\bf v} - {\bf v}'|^2 \; {\bf I} 
- ( {\bf v} - {\bf v}') ( {\bf v} - {\bf v}')}{| {\bf v} - {\bf v}'|^3}
,
\end{equation}
and $\ln \Lambda$ is the Coulomb logarithm. 
The particle mass and charge are denoted by $m_s$ and $e_s$, respectively, 
where the particle species is denoted by the subscript $s ( = a, b )$. 
The distribution function $f_s ({\bf v})$ generally 
depends not only on the 
velocity ${\bf v}$ but also on the position and time variables 
$({\bf x}, t)$ 
although the dependence on $({\bf x}, t)$ are not explicitly shown here.
Writing the distribution function by the sum of  
the equilibrium part and the small perturbation part 
as $f_s = f_{s0} + \delta f_s$, 
we obtain 
\begin{eqnarray}
\label{C012}
C_{ab}  (f_a, f_b) 
& = & 
C_{ab}  (f_{a0}, f_{b0})  
+ C_{ab}  ( \delta f_a, f_{b0})  
\nonumber \\ & & 
\mbox{}
+ C_{ab}  (f_{a0},  \delta f_b )  
+ C_{ab}  ( \delta f_a,  \delta f_b)  
, 
\end{eqnarray}
where the last term $C_{ab}  ( \delta f_a,  \delta f_b)$
is neglected hereafter.   

  We now assume the equilibrium distribution functions to take 
the Maxwellian form 
$f_{s0} = f_{sM} \equiv 
( n_s / \pi^{3/2} v_{Ts}^3) \exp ( - v^2 / v_{Ts}^2 )$  
where $n_s$ is the density, $v_{Ts} \equiv (2T_s/m_s)^{1/2}$ is 
the thermal velocity, and $T_s$ is the temperature. 
   Then, the first term on the right-hand side of Eq.~(\ref{C012}) 
is written as 
\begin{eqnarray}
\label{CabM}
C_{ab}  (f_{aM}, f_{bM}) 
& = & 
- 3 \sqrt{\pi}  \left( \frac{T_a}{T_b}  -  1 \right) 
\frac{f_{aM}}{\tau_{ab}}
x_a
\nonumber \\ & & \mbox{} \times
\left[ G(\alpha_{ab} x_a ) 
- \frac{\alpha_{ab}}{2 x_a} \Phi' (\alpha_{ab} x_a ) \right]
, 
\hspace*{5mm}
\end{eqnarray}
where 
$x_a \equiv v/v_{Ta}$, 
$\alpha_{ab} \equiv v_{Ta}/v_{Tb}$, 
$G(x) \equiv [\Phi(x) - x \Phi'(x)]/(2 x^2)$, 
$\Phi(x) \equiv 2 \pi^{-1/2} \int_0^x e^{-t^2} dt$, and 
$\Phi'(x) \equiv 2 \pi^{-1/2} e^{-x^2} $.  
The collision time $\tau_{ab}$ is defined by 
$(3 \sqrt{\pi}/4) \tau_{ab}^{-1} 
\equiv 4 \pi n_b e_a^2 e_b^2 \ln \Lambda/(m_a^2 v_{Ta}^3)$. 
It is easily seen that $C_{ab}  (f_{aM}, f_{bM})$ vanishes for $T_a = T_b$.  
We hereafter assume that $T_a/T_b = {\cal O}(1)$. 
When $m_a/m_b = {\cal O} (1)$, we have $\alpha_{ab} = {\cal O} (1)$ and 
$C_{ab}  (f_{aM}, f_{bM})  \sim  -  ( T_a / T_b   -  1 ) f_{aM} / \tau_{ab}$. 
In this case, as explained in Appendix~A, 
we may consider that collisions cause species $a$ and $b$ to 
have the equal temperature $T_a = T_b$ 
after a time scale longer than $\tau_{ab}$. 

The second and third terms 
on the right-hand side of Eq.~(\ref{C012}) 
are called the test and field particle parts, respectively, 
and the sum of them gives 
the linearized collision operator,  
\begin{eqnarray}
\label{LCO}
C_{ab}^L (\delta f_a, \delta f_b)
& \equiv & 
C_{ab}(\delta f_a, f_{bM}) + C_{ab} (f_{aM}, \delta f_b)
\nonumber \\ 
& \equiv & 
C_{ab}^T(\delta f_a) + C_{ab}^F (\delta f_b)
. 
\end{eqnarray}
   We now expand the perturbed distribution 
functions $\delta f_s$ $(s=a, b)$ as
\begin{eqnarray}
\label{expansion}
\delta f_s({\bf v}) 
& = & 
\sum_{l=0}^\infty  \delta f_s^{(l)} ({\bf v})
, 
\nonumber \\ 
 \delta f_s^{(l)} ({\bf v})
& = & 
\sum_{m= -l}^{l}
(\delta f_s)_l^m (v) Y_l^m( \theta, \varphi )
, 
\end{eqnarray}
where $Y_l^m ( \theta, \varphi )$ represent spherical harmonic 
functions and $(v, \theta, \varphi)$ are spherical coordinates 
in the velocity space. 
   The $l=1$ component $\delta f_s^{(l=1)}$ of the 
distribution function $\delta f_s$ is further expanded in terms of the 
Laguerre polynomials 
$L_j^{(3/2)}(x_s^2)$ 
$[ L_0^{(3/2)}(x_s^2)=1,L_1^{(3/2)}(x_s^2)=\frac{5}{2}-x_s^2,\cdots ]$ 
as 
\begin{eqnarray} 
\label{l=1}
\delta f_s^{(l=1)} & = & 
f_{sM} \frac{m_s}{T_s} {\bf v} \cdot  
\left[ {\bf u}_s + \frac{2}{5}\frac{{\bf q}_s}{p_s}
\left(x_s^2-\frac{5}{2}\right)
+ \cdots \right] 
\nonumber  \\ & = & 
f_{sM} \frac{m_s}{T_s} {\bf v} \cdot  
\sum_{j=0}^{\infty} {\bf u}_{s j} L_j^{(3/2)}(x_s^2)
,
\end{eqnarray}
where  
$x_s \equiv v / v_{Ts}$. 
   The flow vectors  ${\bf u}_{s j}$ $(j=0, 1, 2, \cdots)$ are defined by 
\begin{eqnarray}
\label{us}
{\bf u}_{s j} & \equiv & 
 \frac{c_j}{n_s}
\int d^3 v \, \delta f_s L_j^{(3/2)}(x_s^2) {\bf v}
,
\nonumber \\
c_j & \equiv & 
\frac{3 \cdot 2^j \cdot j !}{( 2 j + 3 )!!} .
\end{eqnarray}
For $j = 0$ and $j=1$, 
we can write
${\bf u}_{s 0} = {\bf u}_s$ and 
${\bf u}_{s 1} = -(2/5) ({\bf q}_s / p_s)$, 
where ${\bf u}_s \equiv 
n_s^{-1} \int d^3 v\; \delta f_s {\bf v}$ 
and ${\bf q}_s \equiv T_s 
\int d^3 v\; \delta f_s {\bf v} (x_s^2 -\frac{5}{2})$ 
represent 
the fluid velocity and the heat flow, respectively. 

We next use the Laguerre polynomials to expand  the $l=1$ spherical harmonic component of the collision operator before deriving the friction-flow relations 
in Eq.~(\ref{Fai}). 
The resultant expansion [given below in Eq.~(\ref{CabLl=1})] 
contains the coefficients 
(denoted by ${\bf C}_{ab j}$)  
as functionals of  
distribution functions, into which the expression in 
Eq.~(\ref{l=1}) is substituted  
to define the matrix elements $M_{ab}^{jk}$ and 
$N_{ab}^{jk}$ for representing  
the the friction coefficients $l_{ij}^{ab}$ later. 
The $l=1$ component of the collision term 
in Eq.~(\ref{LCO}) 
is written as 
\begin{eqnarray} 
\label{CabLl=1}
C_{ab}^L (\delta f_a^{(l=1)}, \delta f_b^{(l=1)})
& \equiv & C_{ab}^T (\delta f_a^{(l=1)}) 
+ C_{ab}^F(\delta f_b^{(l=1)})
\nonumber \\ 
 & = & 
f_{aM} \frac{m_a}{T_a} {\bf v} \cdot  
\sum_{j=0}^{\infty} {\bf C}_{ab j} L_j^{(3/2)}(x_a^2)
. 
\hspace*{5mm}
\end{eqnarray}
Here, ${\bf C}_{ab j}$ $(j=0,1,2, \cdots)$ are defined by 
\begin{eqnarray}
\label{Caj}
{\bf C}_{ab j} 
& \equiv &
 \frac{c_j}{n_a}
\int d^3 v \; {\bf v} L_j^{(3/2)}(x_a^2) 
C_{ab}^L (\delta f_a, \delta f_b)
\nonumber \\
& = &  
\frac{c_j}{\tau_{ab}} 
\sum_{k=0}^{\infty} 
\left( 
M_{ab}^{jk} {\bf u}_{a k}
+
N_{ab}^{jk} {\bf u}_{b k}
\right)
, 
\end{eqnarray}
where the matrix elements $M_{ab}^{jk}$ and 
$N_{ab}^{jk}$ $(j, k = 0, 1, 2, \cdots)$ are 
given from the test and 
field particle operators, respectively, as~\cite{H&S}  
\begin{eqnarray}
\label{MN}
\frac{n_a}{\tau_{ab}} 
M_{ab}^{jk}
&
\equiv  
&
\int d^3 v \; v_\parallel L_j^{(3/2)} (x_a^2) 
C_{ab}^T 
\left( 
f_{aM} L_k^{(3/2)} (x_a^2) 
\frac{m_a v_\parallel }{T_a} 
\right)
,
\nonumber \\
\frac{n_a}{\tau_{ab}} 
N_{ab}^{jk} 
&
\equiv  
&
\int d^3 v \; v_\parallel L_j^{(3/2)} (x_a^2) 
C_{ab}^F
\left(  
f_{bM} L_k^{(3/2)} (x_b^2)  
 \frac{m_b v_\parallel }{T_b}
\right)
.
\nonumber \\
& & 
\end{eqnarray}
In Eq.~(\ref{MN}), $v_\parallel$ denotes the velocity 
component parallel to the background magnetic field 
although it can be replaced with the velocity component 
in any other direction because of the spherical symmetry 
of the collision operator.

Using the linear collision operator, 
the friction forces ${\bf F}_{ai}$ $(i=1,2,\cdots)$ are given by~\cite{H&S} 
\begin{eqnarray}
\label{Fai}
{\bf F}_{ai}
& \equiv & 
(-1)^{i-1} \int d^3 v \; m_a {\bf v} L^{(3/2)}_{i-1} (x_a^2) 
\sum_b 
C_{ab}^L (\delta f_a, \delta f_b)
\nonumber \\ 
& = & 
(-1)^{i-1} \frac{n_a m_a}{c_{i-1}} \sum_b {\bf C}_{ab, i-1}
\nonumber \\ 
& = & 
(-1)^{i-1}
\sum_b \sum_{j=1}^\infty 
l^{ab}_{ij} {\bf u}_{b, j-1}
\hspace*{2mm}
(i = 1, 2, \cdots)
.
\end{eqnarray}
Here, the first two-order friction forces are 
written as 
${\bf F}_{a 1} = n_a m_a \sum_b {\bf C}_{ab 0}
= \int d^3 v \; m_a {\bf v}\sum_b 
$ 
and
${\bf F}_{a 2} = - \frac{5}{2} n_a m_a {\sum_b} {\bf C}_{ab 1}
= \int d^3 v \; m_a {\bf v} 
\left(x_a^2-\frac{5}{2}\right) \sum_b  
$. 
     The friction coefficients 
$l_{ij}^{ab}$ $(i, j = 1, 2, \cdots)$ are defined by~\cite{H&S}
\begin{equation}
\label{lij}
l^{ab}_{ij}
\equiv 
n_a m_a \left[
\left(
\sum_c
\frac{M_{ac}^{i-1, j-1}}{\tau_{ac}} 
\right)
\delta_{ab} 
+
\frac{ N_{ab}^{i-1, j-1}}{\tau_{ab}}
\right]
,
\end{equation}
where $\delta_{ab}$ denotes the  Kronecker delta 
($\delta_{ab}= 1$ for $a=b$ and $\delta_{ab}= 0$ for $a\neq b$). 

From the momentum conservation in collisions 
[see Eq.~(\ref{momentum}) in Appendix~B], 
we obtain   
\begin{equation}
\label{MN0j}
M_{ab}^{0j} + 
\frac{T_a v_{Ta}}{T_b v_{Tb}} N_{ba}^{0j}
= 0
\hspace*{3mm} ( j = 0, 1, 2, \cdots),
\end{equation}
and 
\begin{equation}
\sum_a l_{1j}^{ab} = 0
\hspace*{3mm} 
( j = 1, 2, \cdots). 
\end{equation}

The adjointness relations for the linearized Landau collision operator 
is written as
\begin{eqnarray}
\label{adjoint}
\int d^3 v \; \frac{\delta f_a}{f_{aM}} C_{ab}^T (\delta g_a) 
& = & 
\int d^3 v \; \frac{\delta g_a}{f_{aM}} C_{ab}^T (\delta f_a)
,
\nonumber \\
T_a 
\int d^3 v \; \frac{\delta f_a}{f_{aM}} C_{ab}^F (\delta f_b) 
& = & 
T_b
\int d^3 v \; \frac{\delta f_b}{f_{bM}} C_{ba}^F (\delta f_a) 
. 
\hspace*{5mm}
\end{eqnarray}
Strictly speaking, 
the linearized Landau operator satisfies 
the adjointness relations in Eq.~(\ref{adjoint}) 
rigorously only for the  case of $T_a = T_b$.  
In this case, 
the symmetry properties of  
$M_{ab}^{ij}$, $N_{ab}^{ij}$, and $l^{ab}_{ij}$ 
are derived from Eq.~(\ref{adjoint})  as 
\begin{eqnarray}
\label{MN_sym}
& & 
M_{ab}^{ij}  =  M_{ab}^{ji}, 
\; \; 
\frac{N_{ab}^{ij}}{T_a v_{Ta}}  =  
\frac{N_{ba}^{ji}}{T_b v_{Tb}} 
\; \; 
(i, j = 0, 1, 2, \cdots)
,
\nonumber \\
& & 
l_{ij}^{ab}=l_{ji}^{ba}
\; \; 
(i, j = 1, 2, \cdots)
. 
\end{eqnarray}
As explained in Appendix~A, 
$T_a$ and $T_b$ are significantly different from each other only when 
$m_a/m_b \ll 1$ or $m_a/m_b \gg 1$. 
It is explained in Appendix~B that, even for this case of unequal temperatures, 
the adjointness relations given in Eq.~(\ref{adjoint}), the symmetry properties 
in Eq.~(\ref{MN_sym}), and Boltzmann's H-theorem 
in the form shown later in Eq.~(\ref{H}) are regarded as 
approximately valid because of the large difference 
between $m_a$ and $m_b$. 
We also should note that 
the Onsager symmetry for collisional 
transport coefficients  
is derived from the adjointness relations which are also used 
to give useful methods for solving drift kinetic equations 
and evaluating neoclassical transport 
fluxes.~\cite{RHH,Hinton,H&S,Balescu,Helander,DKES,Taguchi,Sugama1996,Sugama-Nishimura}

\section{SUGAMA OPERATOR}

The linearized model collision operator for collisions between species 
$a$ and $b$ given by Sugama {\it et al}.~\cite{Sugama2009} is written here as 
\begin{equation}
C_{ab}^{LS}(\delta f_a, \delta f_b )
= C_{ab}^{TS} (\delta f_a) + C_{ab}^{FS} (\delta f_b)
.  
\end{equation}
The test-particle part $C_{ab}^{TS}(\delta f_a)$ 
of the Sugama operator is defined by 
\begin{eqnarray}
\label{CTS}
C_{ab}^{TS} (\delta f_a) 
& = & 
{\cal Q}_{ab} \;
C_{ab}^{T0} \; 
{\cal Q}_{ab} \; \delta f_a 
, 
\nonumber \\ & & 
\end{eqnarray}
where $C_{ab}^{T0}$ is defined by Eq.~(\ref{CT0NF0N}) in Appendix~B 
and the operator ${\cal Q}_{ab}$ is given for an arbitrary distribution 
function $g$ by
\begin{equation}
\label{Qab} 
{\cal Q}_{ab} \; g 
 \equiv g + (\theta_{ab} - 1) 
({\cal P}_{1a} \; g + {\cal P}_{2a} \; g )
,
\end{equation}
with the dimensionless parameter $\theta_{ab}$, 
\begin{equation}
\label{theta_ab}
\theta_{ab} \equiv 
\left[
\frac{T_a \left( \frac{1}{m_a} + \frac{1}{m_b} \right)
}{\left( \frac{T_a}{m_a} + \frac{T_b}{m_b} \right)}
\right]^{1/2}
= 
\left(
\frac{ T_a/T_b  + \alpha_{ab}^2 }{1+ \alpha_{ab}^2}
\right)^{1/2}
.
\end{equation}
The projection operators ${\cal P}_{1a}$ and  ${\cal P}_{2a}$ 
is defined by
\begin{eqnarray}
\label{Pa} 
{\cal P}_{1a} \; g 
& \equiv &
 f_{aM} \frac{m_a}{T_a} 
{\bf u}_a[g] \cdot {\bf v}
,
\nonumber \\ 
{\cal P}_{2a} \; g 
& \equiv &
 f_{aM} \frac{\delta T_a[g]}{T_a} 
\left( x_a^2 - \frac{3}{2} \right) 
,
\end{eqnarray}
where 
${\bf u}_a [g] \equiv n_a^{-1} \int d^3 v \, g {\bf v}$ 
and  
$\delta T_a [g] / T_a
\equiv n_a^{-1} \int d^3 v \, g (m_a v^2/3 T_a - 1 )$.  
   The definition of $\theta_{ab}$ is given so as to 
satisfy
$\int d^3 v \, m_a {\bf v} \, C_{ab}^{TS} (f_{aM} m_a {\bf v}/T_a)
= \int d^3 v \, m_a {\bf v} \, C_{ab}^T (f_{aM} m_a {\bf v}/T_a)$ 
where $C_{ab}^T$ represents the test particle part of the 
linearized Landau collision operator given in Sec.~II. 
 We here note that
$C_{ab}^{TS}$ is defined such that  
the self-adjointness condition, 
\begin{equation}
\label{CTSadjoint}
\int d^3 v \frac{\delta f_a}{f_{aM}} C_{ab}^{TS} (\delta g_a) 
= \int d^3 v \frac{\delta g_a}{f_{aM}} C_{ab}^{TS} (\delta f_a)
\end{equation}
holds exactly even if $T_a \neq T_b$.

The field particle part $C_{ab}^{FS}(\delta f_b)$ of the Sugama 
operator is given by 
\begin{eqnarray}
\label{CFS}
C_{ab}^{FS} (\delta f_b) 
& = & 
- {\bf V}_{ab} [\delta f_b] 
\cdot C_{ab}^{TS} (f_{aM} m_a {\bf v}/T_a)
\nonumber \\ 
&  & \mbox{} 
- W_{ab}[\delta f_b]
 C_{ab}^{TS} (f_{aM} x_a^2)
,
\end{eqnarray}
where
\begin{eqnarray}
\label{alpha_ab}
{\bf V}_{ab} [\delta f_b] 
& \equiv  &
\frac{T_b}{\gamma_{ab}}
\int d^3 v \frac{\delta f_b}{f_{bM}}
C_{ba}^{TS} (f_{bM} m_b {\bf v}/T_b)
,
\end{eqnarray}
and 
\begin{eqnarray}
\label{beta_ab}
W_{ab} [\delta f_b] 
& \equiv  &
\frac{T_b}{\eta_{ab}}
\int d^3 v \frac{\delta f_b}{f_{bM}}
C_{ba}^{TS} (f_{bM} x_b^2)
.
\end{eqnarray}
In Eqs.~(\ref{alpha_ab})--(\ref{beta_ab}), 
\begin{eqnarray}
\label{gamma}
\gamma_{ab}
& \equiv &
T_a
\int d^3 v 
(m_a v_\parallel/ T_a)
C_{ab}^{TS} (f_{aM} m_a v_\parallel /T_a)
\nonumber 
\\ 
& = & 
- \frac{n_a m_a}{\tau_{ab}}
\frac{\alpha_{ab}}{(1+\alpha_{ab}^2)^{3/2}}
\left( \frac{T_a}{T_b} + \alpha_{ab}^2 \right)
\nonumber \\ 
& = & 
-\frac{16\sqrt{\pi}}{3}
\frac{n_a n_b e_a^2 e_b^2 \ln \Lambda}{
(v_{Ta}^2 + v_{Tb}^2)^{3/2}}
\left( \frac{1}{m_a} + \frac{1}{m_b} \right)
,
\end{eqnarray}
and 
\begin{eqnarray}
\label{eta}
\eta_{ab}
& \equiv &
T_a
\int d^3 v 
x_a^2
C_{ab}^{TS} (f_{aM} x_a^2)
\nonumber 
\\ 
& = & 
- \frac{n_a T_a}{\tau_{ab}}
\frac{3 \alpha_{ab}}{(1+\alpha_{ab}^2)^{5/2}}
\left( \frac{T_a}{T_b} + \alpha_{ab}^2 \right)
\nonumber 
\\ 
& = & 
- 8 \sqrt{\pi} \ln \Lambda
\frac{ n_a n_b e_a^2 e_b^2 v_{Ta}^2 v_{Tb}^2}{
(v_{Ta}^2 + v_{Tb}^2)^{5/2}}
\left( \frac{1}{m_a} + \frac{1}{m_b} \right)
\end{eqnarray}
are used. 
We see $\gamma_{ab} = \gamma_{ba}$ and $\eta_{ab} = \eta_{ba}$ 
from Eqs.~(\ref{gamma}) and (\ref{eta}), respectively.
It can be easily verified that the test-particle operator 
$C_{ab}^{TS}$ and the field particle part $C_{ab}^{FS}$ 
defined in Eqs.~(\ref{CTS}) and (\ref{CFS}) 
obey conservation laws for particles, momentum, and  
energy. 
In addition, $C_{ab}^{FS}$ 
satisfies  
the adjointness relation, 
\begin{equation}
\label{CFSadjoint}
T_a \int d^3 v \; \frac{\delta f_a}{f_{aM}} C_{ab}^{FS} (\delta f_b) 
=
T_b \int d^3 v \; \frac{\delta f_b}{f_{bM}} C_{ba}^{FS} (\delta f_a) 
.
\end{equation}

It is shown in Ref.~\cite{Sugama2009} 
that the Sugama operator satisfies 
Boltzmann's H-theorem,
\begin{eqnarray}
\label{H}
T_a \int d^3 v \;   \frac{\delta f_a}{f_{aM}} [ C_{ab}^{TS} (\delta f_a) 
+ C_{ab}^{FS} (\delta f_b) ]  
& &
\nonumber \\ 
+ T_b \int d^3 v \;  \frac{\delta f_b}{f_{bM}} [ C_{ba}^{TS} (\delta f_b) 
+ C_{ba}^{FS} (\delta f_a) ]  
& \leq & 
0
.
\end{eqnarray}
We also find that,  for the case of $m_a/m_b \ll 1$, 
$C_{ab}^{TS}$ and $C_{ab}^{FS}$ coincide with 
$C_{ab}^T$ and $C_{ab}^F$ of the linearized Landau collision operator 
to the lowest order in $(m_a/m_b)^{1/2}$. 
For the case of  $m_a/m_b \gg 1$, 
$C_{ab}^{FS}$ equals $C_{ab}^F$ 
to the lowest order in $(m_b/m_a)^{1/2}$ while, 
for $T_a \neq T_b$,  
$C_{ab}^{TS}$ differs from $C_{ab}^T$ by the 
non-self-adjoint part which remains to the lowest order 
[see the paragraph including Eq.~(\ref{Msymmetry}) in Appendix~B]. 

The matrix elements $M_{ab}^{(S)jk}$ and $N_{ab}^{(S)jk}$ 
$(j,k=0,1,2,\cdots)$ are defined by replacing $C_{ab}^T$ and $C_{ab}^F$ 
with $C_{ab}^{TS}$ and $C_{ab}^{FS}$, respectively, 
in Eq.~(\ref{MN}). 
Similarly, the friction coefficients $l_{ij}^{(S)ab}$ $(i,j=1,2,\cdots)$ 
are defined using $M_{ab}^{(S)jk}$ and $N_{ab}^{(S)jk}$ in Eq.~(\ref{lij}). 
Then, from the momentum conservation law satisfied 
by $C_{ab}^{TS}$ and $C_{ab}^{FS}$, 
we obtain 
\begin{equation}
\label{MS0j}
M_{ab}^{(S)0j} + 
\frac{T_a v_{Ta}}{T_b v_{Tb}} N_{ba}^{(S)0j}
= 0
\hspace*{3mm} ( j = 0, 1, 2, \cdots),
\end{equation}
and 
\begin{equation}
\sum_a l_{1j}^{(S)ab} = 0
\hspace*{3mm} 
( j = 1, 2, \cdots). 
\end{equation}
We also have 
\begin{eqnarray} 
\label{MN00}
& & M_{ab}^{00} 
= - N_{ab}^{00} 
= M_{ab}^{(S)00} 
= - N_{ab}^{(S)00} 
, 
\nonumber \\ 
& & 
 l_{11}^{ab} =  l_{11}^{(S)ab}. 
\end{eqnarray}
From the adjointness relations shown 
in Eqs.~(\ref{CTSadjoint}) and (\ref{CFSadjoint}), 
the symmetry properties of the matrix elements and the friction 
coefficients are derived as
\begin{eqnarray}
\label{MNLS}
& & 
M_{ab}^{(S)ij}  =  M_{ab}^{(S)ji}, 
\; \; 
\frac{N_{ab}^{(S)ij}}{T_a v_{Ta}}  =  
\frac{N_{ba}^{(S)ji}}{T_b v_{Tb}} 
\; \; 
(i, j = 0, 1, 2, \cdots)
, 
\nonumber \\
& & 
l_{ij}^{(S)ab}=l_{ji}^{(S)ba}
\; \; 
(i, j = 1, 2, \cdots)
. 
\end{eqnarray}
In addition, it is found from replacing $(C_{ab}^T, C_{ab}^F)$ with 
$(C_{ab}^{TS}, C_{ab}^{FS})$ in Eq.~(\ref{MN}) and 
using Eqs.~(\ref{CFS}), (\ref{MS0j}), (\ref{MN00}), and 
(\ref{MNLS}) that 
$N_{ab}^{(S)ij}$ $(i,j =0,1,2,\cdots)$ are represented by  
\begin{equation}
\label{NabSij}
N_{ab}^{(S)ij} = 
\frac{M_{ab}^{(S)0i} N_{ab}^{(S)0j}}{M_{ab}^{00}}
= 
\frac{N_{ab}^{(S)i0} N_{ab}^{(S)0j}}{N_{ab}^{00}}
. 
\end{equation}
In Appendix~C, we find 
the detailed expressions of the matrix elements 
$(M_{ab}^{(S)ij}, N_{ab}^{(S)ij})$ as well as $(M_{ab}^{ij}, N_{ab}^{ij})$. 

In the moment method,~\cite{H&S,Sugama-Nishimura} 
the neoclassical transport coefficients, with which 
the radial particle and heat fluxes and the 
parallel current are linearly related to 
the radial density and temperature 
gradients and the parallel electric field, 
can be expressed in terms of the viscosity coefficients and 
the friction coefficients. 
   The friction coefficients $l^{(S) ab}_{ij}$ 
derived from the Sugama operator 
do not all coincide with $l^{ab}_{ij}$ given by 
the Landau operator even for the case of $T_a = T_b$ 
(see Sec.~III.A below). 
   The dependence of the neoclassical transport coefficients 
on the friction coefficients becomes relatively strong 
in the highly collisional regime where accurate values of the 
friction coefficients need to be derived from the model 
collision operator for correctly describing the neoclassical 
transport. 
   In Sec.~IV, the improved Sugama operator is presented 
to reproduce such accurate friction coefficients.

\subsection{Equal temperature case}

   When $T_a = T_b$, 
the test particle part of the Sugama operator is
equivalent to that of the Linearized Landau collision operator,
\begin{equation}
\label{CTSL}
C_{ab}^{TS} = C_{ab}^{T} 
, 
\end{equation}
which can be easily verified from Eq.~(\ref{CTS}) with 
$\theta_{ab} = 1$, ${\cal Q}_{ab}(g) = g$, and 
$C_{ab}^T = C_{ab}^{T0}$ for that case. 
In this equal temperature case, we have 
\begin{eqnarray}
\label{MS}
M_{ab}^{(S) ij} 
& = & 
M_{ab}^{ij} 
, 
\nonumber \\
N_{ab}^{(S) ij} 
& = & 
\frac{N_{ab}^{i0} N_{ab}^{0j}}{N_{ab}^{00}}
\hspace*{3mm} 
(i,j = 0, 1, 2, \cdots), 
\end{eqnarray}
from which 
we see
\begin{eqnarray}
\label{NS}
N_{ab}^{(S) i0} 
& = & 
N_{ab}^{i0}  
\hspace*{3mm} 
(i = 0, 1, 2, \cdots),
\nonumber \\ 
N_{ab}^{(S) 0j} 
& = & 
N_{ab}^{0j}  
\hspace*{3mm} 
(j = 0, 1, 2, \cdots), 
\end{eqnarray}
and 
\begin{eqnarray}
\label{lS}
l^{(S) ab}_{i1} 
& = & 
l^{ab}_{i1}  
\hspace*{3mm} 
(i = 1, 2, \cdots),
\nonumber \\ 
l^{(S) ab}_{1j} 
& = & 
l^{ab}_{1j}  
\hspace*{3mm} 
(j = 1, 2, \cdots).
\end{eqnarray}
We find from Eqs.~(\ref{MN00}) and (\ref{lS}) 
that
the deviations of the friction coefficients 
$l^{(S) ab}_{ij}$ evaluated by the Sugama operator 
from $l^{ab}_{ij}$ by the Landau operator appear 
only for $i \geq 2$ and $j \geq 2$. 
It is also noted that, 
for collisions between particles of like species $(a=b)$, 
the Sugama operator is equivalent to the linearized 
model collision operator 
given in Refs.~\cite{Lin1995,Wang,Abel}.

\section{IMPROVED SUGAMA OPERATOR}

The improved Sugama operator $C_{ab}^{LS{\rm (imp)}}$ 
is defined by adding the correction part $\Delta C_{ab}^{LS}$ 
to the original one $C_{ab}^{LS}$, 
\begin{equation}
\label{imp}
C_{ab}^{LS{\rm (imp)}} (\delta f_a, \delta f_b)
\equiv 
C_{ab}^{LS} (\delta f_a, \delta f_b)
+  \Delta C_{ab}^{LS} (\delta f_a, \delta f_b)
.
\end{equation}
In order for $C_{ab}^{LS{\rm (imp)}}$ 
to reproduce the same friction-flow relations and friction 
coefficients as those in Eqs.~(\ref{Fai}) and (\ref{lij}) 
derived from the linearized Landau collision operator $C_{ab}^L$,  
the correction term 
$\Delta C_{ab}^{LS} (\delta f_a, \delta f_b)$ is defined by 
\begin{equation} 
\label{DCab}
 \Delta C_{ab}^{LS} (\delta f_a, \delta f_b)
 \equiv
f_{aM} \frac{m_a}{T_a} {\bf v} \cdot  
\sum_{j=0}^{\infty} \Delta {\bf C}_{ab j}^L [\delta f_a, \delta f_b]
L_j^{(3/2)}(x_a^2)
,
\end{equation}
with 
\begin{eqnarray}
\label{DCabj}
\Delta {\bf C}_{ab j}^L [\delta f_a, \delta f_b] 
& \equiv & 
\frac{c_j}{\tau_{ab}} 
\sum_{k=0}^{\infty} 
\left(
\Delta M_{ab}^{jk} \; {\bf u}_{a k}[\delta f_a]
\right.
\nonumber \\ & & 
\hspace*{-2mm}
\left.
\mbox{}
+
\Delta N_{ab}^{jk} \; {\bf u}_{b k}[\delta f_b]
\right)
\hspace*{1mm}
(j=0, 1, 2, \cdots)
,
\hspace*{7mm}
\end{eqnarray}
where ${\bf u}_{a k}[\delta f_a]$ and  ${\bf u}_{b k}[\delta f_b]$ 
are evaluated using Eq.~(\ref{us}). 
The corrections $\Delta M_{ab}^{jk}$ and $\Delta N_{ab}^{jk}$ 
of the matrix elements are defined by
\begin{eqnarray}
\Delta M_{ab}^{jk} 
& \equiv & 
M_{ab}^{jk} - M_{ab}^{(S) jk} 
, 
\nonumber \\ 
\Delta N_{ab}^{jk} 
& \equiv & 
N_{ab}^{jk} - N_{ab}^{(S) jk}
\nonumber \\ 
& = & 
\frac{N_{ab}^{00} N_{ab}^{jk} - N_{ab}^{(S)j0} N_{ab}^{(S)0k} }{N_{ab}^{00}} 
,  
\end{eqnarray}
where the matrix elements $M_{ab}^{jk}$ and $N_{ab}^{jk}$ 
($M_{ab}^{(S) jk}$ and  $M_{ab}^{(S) jk}$) are given using 
the test and field particle parts of the Landau operator 
(the original Sugama operator) in Eq.~(\ref{MN}). 
From Eq.~(\ref{MN00}), we immediately find 
\begin{equation}
\Delta M_{ab}^{00} 
=
\Delta N_{ab}^{00} 
= 0
.  
\end{equation}
Using the improved Sugama operator 
$C_{ab}^{LS{\rm (imp)}}$ 
defined by Eqs.~(\ref{imp})--(\ref{DCabj}) instead of 
the linearized Landau collision operator $C_{ab}^L$
to evaluate the matrix elements and friction coefficients in 
Eqs.~(\ref{MN}) and (\ref{lij}), 
we can confirm that $C_{ab}^{LS{\rm (imp)}}$ still 
gives the same values to $M_{ab}^{ij}$, $N_{ab}^{ij}$, and 
$l^{ab}_{ij}$ as $C_{ab}^L$ does, 
and 
accordingly, the improved operator correctly 
reproduces the friction-flow relations in Eq.~(\ref{Fai})  
derived from the Landau operator as well as it retains 
conservation laws of particles, momentum, and energy. 
Therefore, we can expect that the classical and neoclassical transport fluxes 
are accurately evaluated using the improved operator up to the 
highly collisional regime. 
As shown in the literature,~\cite{Balescu,Honda} 
in order to correctly describe the neoclassical transport for the case 
where all particle species belong to the Pfirsch-Schl\"{u}ter collisionality regime, 
we need accurate values for at least the part of the friction coefficients 
$l_{ij}^{ab}$ with $i, j =1,2,3$. 
Accordingly, in this highly collisional case, 
truncation of the summation $\sum_j$ and $\sum_k$ in Eqs.~(\ref{DCab}) 
and (\ref{DCabj}) 
should not be done unless 
the terms with $j \leq 2$ and $k \leq 2$ are retained. 
When the truncation is done such that 
$\Delta M_{ab}^{0k}$ and $\Delta N_{ba}^{0k}$ with 
$k \leq k_{\rm max}$ ($k_{\rm max}$: an arbitrary integer number)
are included, 
the matrix elements associated with the improved operator 
satisfy 
the relations given in the same form as in 
Eq.~(\ref{MN0j}) or Eq.~(\ref{MS0j})
so that the momentum conservation law still holds as well as
the conservation laws of particles and energy. 

We note here that the correction term $\Delta C_{ab}^{LS} (\delta f_a, \delta f_b)$ 
is given for only the $l=1$ spherical harmonic component of the Sugama operator to correctly reproduce the friction-flow relations which are regarded as an important factor in accurate evaluation of collisional transport and flow profiles influencing turbulent transport. 
As pointed out in Ref.~3, 
since the spherical harmonic functions of degree $l$ 
is an eigenfunction of 
the pitch-angle-scattering operator included in the test particle 
collision part with an eigenvalue proportional to $- l ( l +1)$, 
the test particle part tends to be more dominant over  
the field particle part as $l$ is larger. 
Also, in the highly collisional regime, 
anisotropic components of the distribution function represented by 
the spherical harmonic functions of higher $l$'s 
are considered to be stronger damped. 
Thus, without correction terms in the $l \geq 2$ spherical harmonic components of the field particle part, 
the improved operator is expected to 
work accurately for describing the distribution function at high collisionality. 
Besides, in principle, we can extend our procedures to add correction terms to 
all other spherical harmonic components similarly with the approximation 
method of Hirshman and Sigmar.~\cite{HScollision} 
Then, the Landau field particle operator is recovered by using spherical harmonic functions and Laguerre functions of all degrees.

As described in Appendix~B, 
the adjointness relations, the resultant symmetry properties for 
$M_{ab}^{ij}$, $N_{ab}^{ij}$, and $l_{ij}^{ab}$ 
in Eq.~(\ref{MN_sym}), and Boltzmann's H-theorem 
in the form of Eq.~(\ref{H}) are 
not exactly but only approximately satisfied by 
the linearized Landau operator and the improved 
Sugama operator 
for the case of unequal species temperatures 
$T_a \neq T_b$. 
Also, it should be recalled here 
that the two species need to 
have very different masses 
for their temperatures to be significantly different 
from each other.  

When the summations $\sum_j$ and $\sum_k$ in Eqs.~(\ref{DCab}) 
and (\ref{DCabj}) are truncated at the same maximum number 
$j_{\rm max}=k_{\rm max}$, 
the adjointness relations of the improved Sugama operator are still satisfied 
for $T_a = T_b$ 
because the matrix elements $M_{ab}^{jk}$ and $N_{ab}^{jk}$ evaluated by 
the improved operator still keep the symmetry properties.  
On the other hand, the H-theorem is not guaranteed by this truncation 
even for $T_a = T_b$ 
because not all but only some fraction of the 
matrix elements $N_{ab}^{jk}$ $(j,k=0,1,2,\cdots)$ of the Sugama field 
particle operator are replaced with those of the Landau field particle operator. 
[Note that 
the friction-flow relations including 
all matrix elements $M_{ab}^{jk}$ and $N_{ab}^{jk}$ $(j,k=0,1,2,\cdots)$ 
given by the linearized Landau operator is equivalent to the $l=1$ 
spherical harmonic part of that Landau operator which satisfies 
the H-theorem for the $l=1$ parts of the distribution functions.]
It is shown in Ref.~\cite{Sugama2009} that   
the H-theorem for the original Sugama operator can be derived from the fact 
that its field particle part can be completely expressed in terms of 
the test particle part although 
the same technique of the derivation of the H-theorem 
cannot be used for the truncated version of the improved Sugama operator.
However, we can still expect that the H-theorem is approximately 
satisfied by the truncated model if the $l=1$ parts of the 
distribution functions are well represented by the linear 
combinations of only low-order Laguerre polynomials. 
   As shown in Ref.~\cite{Balescu}, sufficiently accurate evaluations of 
collisional (classical and neoclassical) transport fluxes can be made using   
the friction-flow relations including the Laguerre polynomial moments up to the order of $j=2$. 
This appears to be because higher-order Laguerre-polynomial components 
of the distribution functions are 
stronger suppressed by the energy diffusion operator [see Eq.~(\ref{Cv}) in Appendix~B]. 
Therefore, except for the case where the field particle distribution 
takes a special form due to some external sources, 
we don't generally expect that higher $(j\geq 3)$ moments need to be 
retained in Eq.~(41). 

It is easily found from the definition of the improved operator 
in Eqs.~(\ref{imp})--(\ref{DCabj}) 
that 
$C_{ab}^{LS} (\delta f_a, \delta f_b)
= C_{ab}^{LS{\rm (imp)}} (\delta f_a, \delta f_b)$ holds 
 if the perturbed distribution functions 
$\delta f_s$ $(s=a, b)$ include no $l=1$ components 
($\delta f_s^{(l=1)}=0$) [see Eqs.~(\ref{expansion}) 
and (\ref{l=1})]. 
  Therefore,  if $\delta f_s$ $(s=a, b)$ are given by 
the perturbed Maxwellian with 
the perturbed densities $\delta n_s$ and 
temperatures $\delta T_s$ as 
\begin{equation}
\label{shiftM0}
\delta f_s
=
f_{sM}
\left[ 
\frac{\delta n_s}{n_s}
+ \frac{\delta T_s}{T_s}
\left(
\frac{m_s v^2}{2T_s} - \frac{3}{2}
\right)
\right]  
,
\end{equation}
for which $\delta T_a / T_a = \delta T_b/T_b$ is assumed, 
then 
$C_{ab}^{LS{\rm (imp)}} (\delta f_a, \delta f_b)$ vanishes 
as $C_{ab}^{LS} (\delta f_a, \delta f_b)$ 
does.~\cite{Sugama2009} 
   However, 
when $\delta f_s$ $(s=a,b)$ are written as the shifted Maxwellian  
$
\delta f_s
=
f_{sM} (m_a/T_s) ( {\bf u}_s \cdot {\bf v} )
$
with the same flow velocity ${\bf u}_a = {\bf u}_b$ and 
the different equilibrium temperatures $T_a \neq T_b$, 
$C_{ab}^{LS} (\delta f_a, \delta f_b)$ vanishes 
although 
neither $C_{ab}^L (\delta f_a, \delta f_b)$ nor 
$C_{ab}^{LS{\rm (imp)}} (\delta f_a, \delta f_b)$ 
does exactly. 
This is related to the fact that the symmetry properties  
$l_{ab}^{ij} = l_{ba}^{ji}$ $(i, j =1, 2, \cdots)$ are 
slightly broken when $T_a \neq T_b$ 
(see Appendix~D).

When the above-mentioned adjointness relations 
and resultant symmetry properties
are satisfied, 
they provide useful techniques for calculating the neoclassical transport coefficients.~\cite{RHH,Hinton,H&S,Balescu,Helander,DKES,Taguchi,Sugama1996,Sugama-Nishimura}  
Therefore, it will be beneficial for such applications 
if we can have a linearized collision model which satisfies 
the adjoint relations exactly even for $T_a \neq T_b$ while 
giving small inaccuracies to the values of the matrix 
elements and the friction coefficients. 
Such a model is presented in Appendix~D where 
the correction part of the improved Sugama operator 
is symmetrized.

\subsection{Equal temperature case}

When $T_a=T_b$, we use Eqs.~(\ref{MS}) and (\ref{NS}) to obtain 
\begin{eqnarray}
\Delta M_{ab}^{ij} 
& = & 0
, 
\nonumber \\ 
\Delta N_{ab}^{ij} 
& = & 
\frac{N_{ab}^{00} N_{ab}^{ij} - N_{ab}^{i0} N_{ab}^{0j} }{N_{ab}^{00}} 
\hspace*{3mm} 
(i, j = 0, 1, 2, \cdots)
,
\hspace*{3mm} 
\end{eqnarray}
from which we have 
\begin{equation}
\Delta N_{ab}^{00} = \Delta N_{ab}^{i0} = \Delta N_{ab}^{0j} 
= 0
\hspace*{3mm} 
(i, j = 1, 2, \cdots)
.
\end{equation}

\section{COLLISION OPERATOR FOR GYROKINETIC EQUATIONS}

 There are two types of gyrokinetic equations. 
One is the gyrokinetic equation derived by using 
the WKB representation for the perturbed distribution 
function~\cite{Rutherford,Taylor,Antonsen,CTB,F-C,Sugama1998}
which has a high wavenumber in the direction perpendicular to the 
equilibrium magnetic field  ${\bf B}$. 
The other is derived by using the Lie transform technique to 
properly define the gyrocenter coordinates for description of 
the total distribution 
function.~\cite{B&H,Sugama2000} 
The collision operator for the former type of the gyrokinetic 
equation is considered in this section as well as in the 
literature.~\cite{Catto,Xu,Abel,Sugama2009} 
On the other hand,  several studies have been done to represent 
the collision operator for the latter type in 
the gyrocenter 
coordinates.~\cite{Brizard2004,Madsen,Burby,Sugama2015,Esteve,Hirvijoki,Sugama2017} 

When applying the improved Sugama operator 
to the gyrokinetic equation for the perpendicular 
wavenumber vector ${\bf k}_\perp$,  
the collision operator is transformed into the following form,
\begin{eqnarray}
\label{CGK}
& & 
\oint \frac{d \xi}{2\pi}
e^{i {\bf k}_\perp \cdot \mbox{\boldmath $\rho$}_a}
C_{ab}^{LS{\rm (imp)}}(
e^{-i {\bf k}_\perp \cdot
\mbox{\boldmath $\rho$}_a} h_{a {\bf k}_\perp}, 
e^{-i {\bf k}_\perp \cdot
\mbox{\boldmath $\rho$}_b} h_{b {\bf k}_\perp}
)
\nonumber \\ 
& = & 
\oint \frac{d \xi}{2\pi}
e^{i {\bf k}_\perp \cdot \mbox{\boldmath $\rho$}_a}
C_{ab}^{LS}(
e^{-i {\bf k}_\perp \cdot
\mbox{\boldmath $\rho$}_a} h_{a {\bf k}_\perp}, 
e^{-i {\bf k}_\perp \cdot
\mbox{\boldmath $\rho$}_b} h_{b {\bf k}_\perp}
)
\nonumber \\ 
&  & 
\mbox{} + 
\oint \frac{d \xi}{2\pi}
e^{i {\bf k}_\perp \cdot \mbox{\boldmath $\rho$}_a}
\Delta C_{ab}^{LS}(
e^{-i {\bf k}_\perp \cdot
\mbox{\boldmath $\rho$}_a} h_{a {\bf k}_\perp}, 
e^{-i {\bf k}_\perp \cdot
\mbox{\boldmath $\rho$}_b} h_{b {\bf k}_\perp}
)
,
\nonumber \\ & & 
\end{eqnarray}
where $\mbox{\boldmath $\rho$}_a \equiv ({\bf b} \times {\bf v})/\Omega_a$ 
$({\bf b} \equiv {\bf B}/B, \Omega_a \equiv e_a B/m_a c )$ 
and $\oint d \xi / (2\pi)$ represent the gyroradius vector and the gyrophase average, 
respectively, and $h_{a {\bf k}_\perp}$ is given from the 
nonadiabatic part of the perturbed particle distribution function 
$\delta f_{a {\bf k}_\perp} = - (e \phi_{{\bf k}_\perp}/T_a) 
f_{aM}
+ e^{-i {\bf k}_\perp \cdot \mbox{\boldmath $\rho$}_a} 
h_{a {\bf k}_\perp}$.  
The detailed expression of the first term on the right-hand of 
Eq.~(\ref{CGK}) 
is shown in Ref.~\cite{Sugama2009} while the second term is written as 
\begin{eqnarray}
\label{DCGK0}
& & \Delta C_{ab}^{LS(GK)} 
(\delta f_{a {\bf k}_\perp}, \delta f_{b {\bf k}_\perp})
\nonumber \\  & & 
\hspace*{-3mm}
\equiv 
\oint \frac{d \xi}{2\pi}
e^{i {\bf k}_\perp \cdot \mbox{\boldmath $\rho$}_a}
\Delta C_{ab}^{LS}
(\delta f_{a {\bf k}_\perp}, \delta f_{b {\bf k}_\perp})
\nonumber \\  & & 
\hspace*{-3mm}
\equiv 
\frac{m_a}{T_a} \frac{f_{aM}}{\tau_{ab}}
\sum_{j=0}^\infty  c_j  L_j^{(3/2)} (x_a^2)
\nonumber \\ & & 
\mbox{}
\times 
\sum_{k=0}^\infty 
\left[
\Delta M_{ab}^{jk}
\left\{ 
\overline{u}_{\parallel a k} [ h_{a {\bf k}_\perp} ]
J_{0a} 
v_\parallel 
+ 
\overline{u}_{\perp a k} [ h_{a {\bf k}_\perp} ]
J_{1a} 
v_\perp
\right\}
\right. 
\nonumber \\  & & 
\left. 
\mbox{} 
+
\Delta N_{ab}^{jk}
\left\{ 
\overline{u}_{\parallel b k} [ h_{b {\bf k}_\perp} ]
J_{0a}
v_\parallel 
+ 
\overline{u}_{\perp b k} [ h_{b {\bf k}_\perp} ]
J_{1a}
v_\perp
\right\}
\right]
,
\end{eqnarray}
where $J_{0s} \equiv  J_0 (k_\perp v_\perp/\Omega_s)$ 
and $J_{1s} \equiv  J_1 (k_\perp v_\perp/\Omega_s)$ $(s=a,b)$
denote the zeroth- and first-order Bessel functions of 
the normalized perpendicular wavenumber $k_\perp v_\perp/\Omega_s$, 
respectively, and 
\begin{eqnarray}
\label{DCGK1}
\overline{u}_{\parallel s k} [ h_{s {\bf k}_\perp} ]
& \equiv & 
\frac{c_k}{n_s}
\int d^3 v \;  L_k^{(3/2)} (x_s^2)  h_{s {\bf k}_\perp}
J_{0s}
v_\parallel 
,
\nonumber \\  
\overline{u}_{\perp s k} [ h_{s {\bf k}_\perp} ]
& \equiv & 
\frac{c_k}{n_s}
\int d^3 v \; L_k^{(3/2)} (x_s^2) h_{s{\bf k}_\perp}
J_{1s} 
v_\perp 
, 
\hspace*{3mm}
\end{eqnarray}
are used. 

In the case of application to the drift kinetic equation 
for studying neoclassical transport, 
we neglect the finite gyroradius effects and 
take the limit 
${\bf k}_\perp \cdot \mbox{\boldmath $\rho$}_s 
\rightarrow 0$ $(s=a, b)$. 
Then, we put   
$J_{0s} \rightarrow 1$, $J_{1s} \rightarrow 0$, 
and $\overline{u}_{\perp s k} [ h_{s {\bf k}_\perp} ] \rightarrow 0$ 
in Eqs.~(\ref{DCGK0}) and (\ref{DCGK1}).

\section{CONCLUSIONS}

In this paper, the improved linearized model collision operator 
which can be applied up to the highly collisional regime 
is presented. 
The improved operator is constructed by adding the correction part to 
the previous model by Sugama {\it et al}.\ 
so as to reproduce the same friction-flow relations as those given by 
the linearized Landau collision operator. 
In the improved model, 
conservation laws of particles, momentum, and energy are retained 
while the adjointness relations and Boltzmann's H-theorem are 
approximately valid for collisions between unlike particle species with 
unequal temperatures and very different masses. 
It is also shown that the improved operator can be modified to 
satisfy the adjointness relations exactly even in the unequal-temperature 
case. 
This modification causes the friction coefficients to deviate 
from those given by 
the Landau operator although the influence of 
the deviations is made small by the very different masses. 

Performing the gyrophase average
with keeping the finite gyroradius effect, 
the improved operator is represented in the suitable 
form for gyrokinetic equations. 
In the zero-gyroradius limit, 
the gyrophase-averaged improved 
operator can be used in drift kinetic equations 
to accurately evaluate neoclassical transport in all collisionality regimes. 
It is considered that only the terms with $j \leq 2$ in the Laguerre polynomial 
expansion of the correction part of the operator need to be kept 
even for the most collisional case where all particle species are in the 
Pfirsch-Schl\"{u}ter collisionality regime. 
The present model is expected to be useful for simulation studies of
neoclassical and turbulent transport processes in plasmas including 
multi-species of particles in various collisional regimes.

\begin{acknowledgments}
The authors sincerely thank Dr.\ Y. Idomura and Dr.\ K. Obrejan 
for beneficial discussion on kinetic simulation of impurity transport.  
This work is supported in part by 
JSPS Grants-in-Aid for Scientific Research Grant No.~19H01879 
 and in part by the NIFS Collaborative Research Program NIFS18KNTT045. 
\end{acknowledgments}

\appendix

\section{COLLISIONAL ENERGY TRANSFER BETWEEN UNLIKE SPECIES WITH UNEQUAL TEMPERATURES}

Using Eq.~(\ref{CabM}), the collisional energy transfer from species $b$ to $a$, 
which equals the opposite sign of that from species $a$ to $b$, is given by 
\begin{eqnarray}
\label{energy_transfer}
& & 
\hspace*{-5mm}
\int d^3 v \; 
C_{ab}  (f_{aM}, f_{bM}) 
\frac{m_a v^2}{2}
=
- \int d^3 v \; 
C_{ba}  (f_{bM}, f_{aM}) 
\frac{m_b v^2}{2}
\nonumber \\
& & =
- 3 
\frac{m_a \alpha_{ab}^3}{m_b ( 1 + \alpha_{ab}^2 )^{3/2}} 
\frac{n_a (T_a - T_b)}{\tau_{ab}} 
, 
\end{eqnarray}
where each species is assumed to be in the local equilibrium state 
represented by the Maxwellian distribution function.  
Thus, 
if $T_a \neq T_b$,  
collisions cause the temperatures of the two species to approach to each other,
and the characteristic rate $\nu_{ab}^{et}$ 
of the collisional energy transfer from species 
$b$ to $a$ is given by 
\begin{eqnarray}
\label{rate}
\nu_{ab}^{et} 
& = & 
\frac{m_a \alpha_{ab}^3}{m_b ( 1 + \alpha_{ab}^2 )^{3/2}} 
\frac{1}{\tau_{ab}} 
=
\frac{e_b^2 n_b m_a \alpha_{ab}^3}{e_a^2 
n_a m_b ( 1 + \alpha_{ab}^2 )^{3/2}} 
\frac{1}{\tau_{aa}} 
\nonumber \\ 
& = & 
\frac{e_a^2 m_b}{e_b^2 m_a ( 1 + \alpha_{ab}^2 )^{3/2}} 
\frac{1}{\tau_{bb}} 
. 
\end{eqnarray}

We now assume that 
$|e_a/e_b| = {\cal O}(1)$, $n_a/n_b = {\cal O}(1)$, 
and $T_a/T_b = {\cal O}(1)$. 
Then, in the case where $m_a/m_b = {\cal O} (1)$, 
we have $\alpha_{ab} = {\cal O} (1)$ and accordingly 
$\nu_{ab}^{et}  \sim 1/ \tau_{aa} \sim 1/ \tau_{bb}$ from Eq.~(\ref{rate}). 
This implies that, 
the relaxation toward the equal-temperature $(T_a = T_b)$ state 
due to the unlike-species collisions 
and the thermalization toward the Maxwellian equilibrium 
are expected to have occurred on the same time scale 
and that the Maxwellian distribution functions $f_{aM}$ 
and $f_{bM}$ should have the same temperature $T_a = T_b$. 

Next, we consider another case where 
$m_a/m_b \ll 1$ or $m_a/m_b \gg1$ holds. 
Then, $\alpha_{ab} \gg 1$ and 
$\nu_{ab}^{et} \sim (m_a/m_b) / \tau_{aa} \sim (m_a/m_b)^{1/2} / \tau_{bb}$ 
are obtained for $m_a/m_b \ll 1$ while 
$\alpha_{ab} \ll 1$ and 
$\nu_{ab}^{et} \sim (m_b/m_a)^{1/2} / \tau_{aa}\sim (m_b/m_a) / \tau_{bb}$ 
for $m_a/m_b \gg 1$. 
Therefore,  when $m_a/m_b \ll 1$ or $m_a/m_b \gg1$, 
collisional energy exchange between species $a$ and $b$ is so slow 
that $f_{aM}$ and $f_{bM}$ are permitted to have unequal temperatures 
$T_a \neq T_b$. 
 
We now consider the case where $m_a/m_b \gg 1$, 
$|e_a/e_b| \gg 1$, $n_a/n_b \ll 1$, 
and $T_a/T_b = {\cal O}(1)$.  
This can happen when $a$ and $b$ represent heavy minority impurity ions with 
high charge number and bulk hydrogen isotopes (or electrons), respectively.  
Then, we have $\alpha_{ab} \ll 1$ and 
\begin{equation}
\label{rate2}
\nu_{ab}^{et} 
\sim 
\frac{e_b^2 n_b  m_b^{1/2}}{e_a^2 n_a m_a^{1/2}} 
\frac{1}{\tau_{aa}} 
\sim 
\frac{e_a^2 m_b}{e_b^2 m_a} 
\frac{1}{\tau_{bb}} 
. 
\end{equation}
%
For example, 
using Eq.~(\ref{rate2}) in realistic cases as in the JET and ASDEX Upgrade experiments shown in Ref.~\cite{Casson}, 
the characteristic rates
$\nu_{Wi}^{et}$ and $\nu_{We}^{et}$ of the collisional energy transfer from tungsten impurities (W) to bulk hydrogen isotopes ($i$) and to electrons ($e$) are 
estimated to satisfy
\begin{equation}
\nu_{Wi}^{et} \sim 1/\tau_{WW},
\hspace*{5mm}
\nu_{Wi}^{et}  >  1/\tau_{ii}, 
\end{equation}
and 
\begin{equation}
\nu_{We}^{et} \ll 1/\tau_{WW} ,
\hspace*{5mm}
\nu_{We}^{et} \ll 1/\tau_{ee} ,
\end{equation}
respectively. 
Under these conditions, it is reasonable to assume $T_W = T_i$ 
although it is not to assume $T_W = T_e$. 

\section{EFFECTS OF UNEQUAL TEMPERATURES ON ADJOINTNESS RELATIONS}

Based on the Landau collision operator defined in Eq.~(\ref{CLandau}), 
the test and field particle parts are written as 
\begin{eqnarray}
\label{C0N}
C_{ab}^T(\delta f_a) 
& \equiv & 
C_{ab}^{T0}(\delta f_a) + C_{ab}^{TN}(\delta f_a) 
,
\nonumber \\
C_{ab}^F (\delta f_b) 
& \equiv & 
C_{ab}^{F0}(\delta f_b) + C_{ab}^{FN}(\delta f_b) 
,  
\end{eqnarray}
where 
\begin{eqnarray}
\label{CT0NF0N}
C_{ab}^{T0}(\delta f_a) 
& \equiv & 
 \frac{2\pi e_a^2 e_b^2 \ln \Lambda}{m_a^2}
\frac{\partial}{\partial {\bf v}}
\cdot \left[ f_{aM} ({\bf v})
\frac{\partial}{\partial {\bf v}}
\left( \frac{\delta  f_a ({\bf v})}{f_{aM} ({\bf v})} \right)
\right. \nonumber \\ & & 
\left. \mbox{} 
\cdot 
\int d^3 v' \; f_{bM} ({\bf v}')
{\bf U} ({\bf v}-{\bf v}' )  
\right]
\nonumber \\ & \equiv & 
\nu_D^{ab} (v) {\cal L} (\delta f_a )
+ {\cal C}_v^{ab} (\delta f_a)
,
\nonumber \\
C_{ab}^{TN}(\delta f_a) 
& \equiv & 
\left( \frac{1}{T_b} - \frac{1}{T_a} \right)
\frac{2\pi e_a^2 e_b^2 \ln \Lambda}{m_a}
\frac{\partial}{\partial {\bf v}}
 \nonumber \\ & & 
\mbox{} 
\cdot \left[ \delta f_a ({\bf v}) {\bf v}
\cdot 
\int d^3 v' \; f_{bM} ({\bf v}')
{\bf U} ({\bf v}-{\bf v}' )  
\right]
\nonumber \\ & \equiv & 
\left( \frac{1}{T_b} - \frac{1}{T_a} \right)
\frac{m_a}{v^2} 
\frac{\partial}{\partial v} 
\left[
\frac{\nu_\parallel^{ab} (v)}{2} v^5
\delta f_a
\right]
,
\nonumber \\
C_{ab}^{F0}(\delta f_b) 
& \equiv & 
 - \frac{2\pi e_a^2 e_b^2 \ln \Lambda}{m_a m_b}
\frac{\partial}{\partial {\bf v}}
\cdot \left[ f_{aM} ({\bf v})
\right. \nonumber \\ & & 
\left. \mbox{} 
\hspace*{-5mm}
\cdot 
\int d^3 v' \; f_{bM} ({\bf v}')
{\bf U} ({\bf v}-{\bf v}' )  
\cdot
\frac{\partial}{\partial {\bf v}'}
\left( \frac{\delta  f_b ({\bf v}')}{f_{bM} ({\bf v}')} \right)
\right]
,
\nonumber \\
 C_{ab}^{FN}(\delta f_b) 
& \equiv & 
\left( \frac{1}{T_b} - \frac{1}{T_a} \right)
\frac{2\pi e_a^2 e_b^2 \ln \Lambda}{m_a}
\frac{\partial}{\partial {\bf v}}
 \nonumber \\ & & 
\hspace*{-3mm}
\mbox{} 
\cdot \left[ f_{aM} ({\bf v}) {\bf v}
\cdot 
\int d^3 v' \; \delta f_b ({\bf v}')
{\bf U} ({\bf v} - {\bf v}' )  
\right]
.  
\end{eqnarray}
Here, $C_{ab}^{T0}(\delta f_a)$ consists of 
the pitch-angle-scattering part 
$\nu_D^{ab} (v) {\cal L} (\delta f_a)$
and the energy diffusion part ${\cal C}_v^{ab} (\delta f_a)$. 
The pitch-angle-scattering operator
${\cal L}$ is defined by 
\begin{eqnarray}
\label{pitch}
{\cal L}(\delta f_a)
 & \equiv & 
\frac{1}{2}
\frac{\partial}{\partial {\bf v}} \cdot
\left[
\left( v^2 {\bf I} - {\bf v} {\bf v} \right)
\cdot \frac{\partial \delta f_a}{\partial {\bf v}}
\right]
\nonumber \\ 
&  = & 
\frac{1}{2}
\left[
\frac{1}{\sin\theta}
\frac{\partial}{\partial\theta}
\left(
\sin\theta \frac{\partial \delta f_a}{\partial\theta}
\right)
+ \frac{1}{\sin^2\theta}
\frac{\partial^2 \delta f_a}{\partial \varphi^2}
\right] 
, 
\hspace*{10mm}
\end{eqnarray}
where ${\bf I}$ denotes the unit tensor 
and $(v, \theta, \varphi)$ represent 
spherical coordinates in the velocity space. 
The energy diffusion operator ${\cal C}_v^{ab}$  
is defined by 
\begin{equation}
\label{Cv}
{\cal C}_v^{ab} ( \delta f_a )
\equiv 
\frac{1}{v^2} 
\frac{\partial}{\partial v} 
\left[
\frac{\nu_\parallel^{ab} (v)}{2} v^4
f_{aM}
\frac{\partial}{\partial v} 
\left(
\frac{\delta f_a}{f_{aM}}
\right)
\right]
.
\end{equation}
The collision frequencies for pitch-angle scattering 
and energy diffusion are given by  
$\nu_D^{ab} (v)\equiv(3 \sqrt{\pi}/4) \tau_{ab}^{-1} 
[\Phi(x_b)-G(x_b)]/x_a^3$
and  
$\nu_\parallel^{ab} (v)\equiv(3 \sqrt{\pi}/2) \tau_{ab}^{-1} 
G(x_b) / x_a^3$, 
respectively, where
$(3 \sqrt{\pi}/4) \tau_{ab}^{-1} 
\equiv 4 \pi n_b e_a^2 e_b^2 \ln \Lambda/(m_a^2 v_{Ta}^3)$
($\ln \Lambda$: The Coulomb logarithm), 
$\Phi(x) \equiv 2 \pi^{-1/2} \int_0^x e^{-t^2} dt$, 
$G(x) \equiv [\Phi(x) - x \Phi'(x)]/(2 x^2)$, 
$x_s \equiv v/v_{Ts}$, and $v_{Ts} \equiv (2T_s/m_s)^{1/2}$ $(s = a, b)$.
We can easily confirm that $\nu_D {\cal L}$, $C_v^{ab}$, 
and accordingly $C_{ab}^{T0}$ are all self-adjoint so that 
\begin{equation}
\label{T0adjoint}
\int d^3 v \; \frac{\delta f_a}{f_{aM}} C_{ab}^{T0} (\delta g_a) 
=
\int d^3 v \; \frac{\delta g_a}{f_{aM}} C_{ab}^{T0} (\delta f_a) 
\end{equation}
holds for arbitrary functions $\delta f_a$ and $\delta g_a$ of ${\bf v}$. 
It can also be shown that $C_{ab}^{F0}$ satisfies the adjointness relation 
written as 
\begin{equation}
\label{F0adjoint}
\int d^3 v \; \frac{\delta f_a}{f_{aM}} C_{ab}^{F0} (\delta f_b) 
=
\int d^3 v \; \frac{\delta f_b}{f_{bM}} C_{ba}^{F0} (\delta f_a) 
.
\end{equation}
The remaining test and field particle operators $C_{ab}^{TN}$ and 
$C_{ab}^{FN}$ do not keep adjoint relations such as   
Eqs.~(\ref{T0adjoint}) and (\ref{F0adjoint}) satisfied by 
$C_{ab}^{T0}$ and $C_{ab}^{F0}$, respectively, although 
$C_{ab}^{TN}$ and $C_{ab}^{FN}$  vanish for $T_a = T_b$. 

We also note that the two pairs of the operators 
$(C_{ab}^{T0}, C_{ab}^{F0})$ 
and $(C_{ab}^{TN}, C_{ab}^{FN})$ independently satisfy 
the particle, momentum and energy conservation laws, which 
are written as  
\begin{equation}
\label{particle}
\int d^3 v \;  C_{ab}^{TA} (\delta f_a) 
= \int d^3 v \;  C_{ab}^{FA} (\delta f_b) =0
\hspace*{5mm} ( A = 0, N ), 
\end{equation}
\begin{eqnarray}
\label{momentum}
& & 
\int d^3 v \;  m_a {\bf v} C_{ab}^{TA} (\delta f_a) 
+ \int d^3 v \;  m_b {\bf v} 
C_{ba}^{FA} (\delta f_a) 
\nonumber \\   & & 
\hspace*{5mm} 
= 0   
\hspace*{5mm} ( A = 0, N ), 
\end{eqnarray}
and  
\begin{eqnarray}
\label{energy}
&  & 
\int d^3 v \;  \frac{1}{2}m_a v^2  C_{ab}^{TA} (\delta f_a) 
+ \int d^3 v \;  \frac{1}{2}m_b v^2 
C_{ba}^{FA} (\delta f_a)   
\nonumber \\   & & 
\hspace*{5mm} 
= 0   
\hspace*{5mm} ( A = 0, N ), 
\end{eqnarray}
respectively. 

From the Galilean invariance and spherical symmetry of the Landau 
collision operator, 
we have an identity,   
$\int d^3 v \; m_a ({\bf v}- {\bf u})
C_{ab} [f_{aM} ({\bf v} - {\bf u}), f_{bM} ({\bf v} - {\bf u}) ]
=
\int d^3 v \; m_a {\bf v}
C_{ab} [f_{aM} ({\bf v}), f_{bM} ({\bf v}) ]=0$, 
for an arbitrary vector ${\bf u}$ which is independent of 
${\bf v}$. 
Then, taking the ${\bf u} \rightarrow 0$ limit of the above identity 
and using the particle and momentum conservation laws, 
we can derive another type of relations,  
\begin{eqnarray}
\label{useful}
& & \int d^3 v \;  m_a {\bf v} \; C_{ab}^T (f_{aM} m_a {\bf v} / T_a) 
\nonumber \\
& =  & 
\int d^3 v \;  m_b {\bf v} \; C_{ba}^T (f_{bM} m_b {\bf v} / T_b)
\nonumber \\
& =  & 
- \int d^3 v \;  m_a {\bf v} \; C_{ab}^F (f_{bM} m_b {\bf v} / T_b) 
\nonumber \\
& =  & 
- \int d^3 v \;  m_b {\bf v} \; C_{ba}^F (f_{aM} m_a {\bf v} / T_a)  
,
\end{eqnarray}
We should note that the symmetry properties shown in 
Eq.~(\ref{useful}) are valid even when $T_a \neq T_b$ although 
they are not satisfied in the same way as 
Eqs.~(\ref{particle})--(\ref{energy}) are separately satisfied 
by the two pairs 
of the operators $(C_{ab}^{TA}, C_{ab}^{FA})$ $(A=0, N)$ 
for $T_a \neq T_b$. 

Using Eq.~(\ref{C0N}), the matrix elements $M_{ab}^{ij}$ and $N_{ab}^{ij}$, 
which are defined by Eq.~(\ref{MN}),  
are written as 
\begin{eqnarray}
\label{MN0N}
M_{ab}^{ij} 
& = & 
M_{ab}^{(0)ij} + M_{ab}^{(N)ij}
,
\nonumber \\
N_{ab}^{ij} 
& = & 
N_{ab}^{(0)ij} + N_{ab}^{(N)ij}
,
\end{eqnarray}
where $M_{ab}^{(A)ij}$ and $N_{ab}^{(A)ij}$ $(A=0,N)$ 
are defined by
\begin{eqnarray}
\label{MN0Nij}
\hspace*{-5mm}
\frac{n_a}{\tau_{ab}} 
M_{ab}^{(A)ij}
&
\equiv  
&
\int d^3 v \; v_\parallel L_i^{(3/2)} (x_a^2) 
C_{ab}^{TA}
\left( 
f_{aM} L_j^{(3/2)} (x_a^2) 
\frac{m_a v_\parallel }{T_a} 
\right)
,
\nonumber \\
\hspace*{-5mm}
\frac{n_a}{\tau_{ab}} 
N_{ab}^{(A)ij} 
&
\equiv  
&
\int d^3 v \; v_\parallel L_i^{(3/2)} (x_a^2) 
C_{ab}^{FA}
\left(  
f_{bM} L_j^{(3/2)} (x_b^2)  
 \frac{m_b v_\parallel }{T_b}
\right).  
\nonumber \\ & & 
\end{eqnarray}
Then, the momentum conservation law shown in 
Eq.~(\ref{momentum}) is used to find   
\begin{equation}
\label{mom_A}
M_{ab}^{(A)0j} + 
\frac{T_a v_{Ta}}{T_b v_{Tb}} N_{ba}^{(A)0j}
= 0
\hspace*{3mm} ( A = 0, N ;  j = 0, 1, 2, \cdots),
\end{equation}
The symmetry properties of  $M_{ab}^{(0)ij}$ and $N_{ab}^{(0)ij}$ 
are derived from the adjointness relations given by 
Eqs.~(\ref{T0adjoint}) and (\ref{F0adjoint})  
as 
\begin{equation}
\label{symmetry0}
M_{ab}^{(0)ij} 
=  M_{ab}^{(0)ji}
\hspace*{3mm}
\mbox{and}
\hspace*{3mm}
\frac{N_{ab}^{(0)ij}}{T_a^2 v_{Ta}}
= 
\frac{N_{ba}^{(0)ji}}{T_b^2 v_{Tb}}
, 
\end{equation}
respectively. 
Also from Eq.~(\ref{useful}), we obtain 
\begin{equation}
\label{symmetry00}
M_{ab}^{00} 
=  - N_{ab}^{00}
,
\hspace*{6mm}
\frac{N_{ab}^{00}}{T_a v_{Ta}}
= 
\frac{N_{ba}^{00}}{T_b v_{Tb}}
.
\end{equation}
It should be noted that the symmetry properties of $N_{ab}^{(0)ij}$ and 
$N_{ab}^{00}$ take different forms with respect to the way the temperatures 
enter. 

     In the case of $m_a/m_b = {\cal O}(1)$, 
the temperatures $T_a$ and $T_b$ are expected to be close to each other 
because of the relatively fast energy exchange due to collisions. 
Therefore, only when  
$m_a \ll m_b$  or $m_a \gg m_b$, 
$T_a$ can be significantly different from $T_b$.   
   In the limiting case $m_a \ll m_b$, it is shown that  
the pitch-angle-scattering term 
$\nu_D(v){\cal L} (\delta f_a)$ is dominant 
in the test particle operator $C_{ab}^T (\delta f_a)$ where  
the energy scattering term $C_v^{ab}(\delta f_a)$ and 
the non-adjoint part $C_{ab}^{TN}(\delta f_a)$ are negligible 
in the lowest order of the expansion with respect to $(m_a/m_b)^{1/2}$. 
However, when $T_a \neq T_b$, 
$C_{ab}^{FN}(\delta f_b)$ is not negligible 
but it is necessary to keep 
contributions from both $C_{ab}^{F0}(\delta f_b)$ 
and $C_{ab}^{FN}(\delta f_b)$ for 
accurately evaluating collisional momentum transfer. 
Then, it can be shown that, to the lowest order in $(m_a/m_b)^{1/2}$,
the test and field particle parts of 
the Sugama operator 
$C_{ab}^{S} (\delta f_a, \delta f_b)
= C_{ab}^{TS} (\delta f_a) + C_{ab}^{FS} (\delta f_b)$ 
correctly approximate  $C_{ab}^T (\delta f_a)$ and 
$C_{ab}^F (\delta f_b)$ of the linearized Landau operator, respectively. 

We next consider the case in which $m_a \gg m_b$ and $T_a \neq T_b$ hold.  
In this case, $C_{ab}^{TN}(\delta f_a)$ is not negligibly small compared with 
$C_{ab}^{T0}(\delta f_a)$  while $C_{ab}^{FN}(\delta f_b)$ does not contribute 
to $C_{ab}^F(\delta f_b)$ in the lowest order of the expansion 
with respect to $(m_b/m_a)^{1/2}$. 
Then, $C_{ab}^F(\delta f_b)$ is well approximated by 
either $C_{ab}^{F0}(\delta f_b)$ or $C_{ab}^{FS} (\delta f_b)$ 
although the difference of $C_{ab}^T(\delta f_a)$ from 
$C_{ab}^{T0}(\delta f_a)$ or $C_{ab}^{TS} (\delta f_a)$ 
is significant. 
However, this difference doesn't cause serious errors in  
solving the kinetic equation for $\delta f_a$ 
as far as $C_{ab}^T(\delta f_a)/C_{aa}^T(\delta f_a) 
\sim (e_b/e_a)^2 (n_b/n_a)  (m_b/m_a)^{1/2}$ becomes very small.  
[This ratio $(e_b/e_a)^2 (n_b/n_a)  (m_b/m_a)^{1/2}$ can be 
large in such a case of tungsten impurity as mentioned in Appendix~A 
although,  for that case,  $T_a =T_b$ is expected so that 
$C_{ab}^T(\delta f_a) = C_{ab}^{T0}(\delta f_a) = C_{ab}^{TS} (\delta f_a)$ 
holds.]
Except for this limiting case of $m_a \gg m_b$ and $T_a \neq T_b$, 
we can suppose that the matrix elements $M_{ab}^{ij}$ evaluated by 
$C_{ab}^T (\delta f_a) = C_{ab}^{T0} (\delta f_a) + C_{ab}^{TN} (\delta f_a)$ 
satisfy the 
symmetry relations of the same form as those for $M_{ab}^{(0)ij}$   
shown in Eq.~(\ref{symmetry0}), 
\begin{equation}
\label{Msymmetry}
M_{ab}^{ij}  =  M_{ab}^{ji}
\; \; 
(i, j = 0, 1, 2, \cdots)
.
\end{equation}
It is recalled that contributions of $C_{ab}^{TN}(\delta f_a)$ to the collisional 
momentum transfer are taken into account in defining  
 $C_{ab}^{TS} (\delta f_a)$ such that $C_{ab}^{TS} (\delta f_a)$ and 
$C_{ab}^T (\delta f_a) = C_{ab}^{T0} (\delta f_a) + C_{ab}^{TN} (\delta f_a)$ 
give the same matrix element $M_{ab}^{00}$ even 
when $m_a \gg m_b$ and $T_a \neq T_b$. 
Also, $C_{ab}^{TS} (\delta f_a)$ 
is constructed so as to yield the matrix elements 
$M_{ab}^{(S)ij}$ which satisfy symmetry relations of the same form as 
in Eq.~(\ref{Msymmetry}).

When $m_a/m_b \ll 1$ and $T_a \neq T_b$, 
 $C_{ab}^{FN}(\delta f_a)$ makes a significant contribution 
to $C_{ab}^F (\delta f_a) = C_{ab}^{F0}(\delta f_a) 
+ C_{ab}^{FN}(\delta f_a)$. 
In this case, we can show that, to the lowest order in $(m_a/m_b)^{1/2}$, 
\begin{equation}
\label{Nsymmetry}
\frac{N_{ab}^{ij}}{T_a v_{Ta}}  =  
\frac{N_{ba}^{ji}}{T_b v_{Tb}} 
\; \; 
(i, j = 0, 1, 2, \cdots)
\end{equation}
are satisfied 
by the matrix elements $N_{ab}^{ij}$ associated 
with $C_{ab}^F (\delta f_a)$. 
Note that the second relation in Eq.~(\ref{symmetry00}), 
which holds exactly,
is included as a special case in 
the symmetry relations shown by Eq.~(\ref{Nsymmetry}) 
and that they take a different form from those for $N_{ab}^{(0)ij}$
in Eq.~(\ref{symmetry0}).  
Also, the matrix elements 
$N_{ab}^{(S)ij}$ evaluated by $C_{ab}^{FS} (\delta f_a)$ 
satisfy symmetry relations of the same form as 
in Eq.~(\ref{Nsymmetry}). 

In summary, 
the adjointness relations of the linearized Landau operator $C_{ab}^L$ 
are not satisfied rigorously in collisions between unlike species 
with unequal temperatures 
although significantly different temperatures occur in the case  
where the two species have so different masses that the 
adjointness relations and symmetry properties of the matrix 
elements and the friction coefficients can still be used as 
approximately valid formulas. 
On the other hand, 
the Sugama operator $C_{ab}^{LS}$ in Sec.~III 
and the operator $C_{ab}^{LS*{\rm (imp)}}$ in Appendix~D 
are constructed so as to exactly keep the adjointness 
relations
which can be useful in formulating efficient methods 
of evaluating Onsager symmetric collisional transport 
coefficients.~\cite{RHH,Hinton,H&S,Balescu,Helander,DKES,Taguchi,Sugama1996,Sugama-Nishimura}

\section{MATRIX ELEMENTS ASSOCIATED WITH THE LINEARIZED LANDAU OPERATOR AND THE SUGAMA OPERATOR}

This Appendix shows how the matrix elements 
$M_{ab}^{ij} = M_{ab}^{(0)ij} + M_{ab}^{(N)ij}$ and 
$M_{ab}^{ij} = M_{ab}^{(0)ij} + M_{ab}^{(N)ij}$ 
[see Eqs.~(\ref{MN0N}) and (\ref{MN0Nij}) in Appendix~B] 
which are associated with the test part 
$C_{ab}^T = C_{ab}^{T0} + C_{ab}^{TN}$ and 
the field part 
$C_{ab}^F = C_{ab}^{F0} + C_{ab}^{FN}$ of 
the linearized Landau operator $C_{ab}^L = C_{ab}^T + C_{ab}^F$ 
are expressed in terms of $\alpha_{ab}\equiv v_{Ta}/v_{Tb}$, 
$T_a/T_b$, and $m_a/m_b$. 
In addition, it is shown how to evaluate
$M_{ab}^{(S)ij}$ and $N_{ab}^{(S)ij}$ defined from 
the Sugama operator 
$C_{ab}^{LS}= C_{ab}^{TS} + C_{ab}^{FS}$ (see Sec.~III). 

First, the $00$ elements of the matrices 
$M_{ab}^{(A) ij}$ and $N_{ab}^{(A) ij}$ $(A=0,N)$ 
are written as follows:
\begin{eqnarray} 
\label{00}
M_{ab}^{(0) 00} 
& = & 
- \frac{\alpha_{ab}}{(1+ \alpha_{ab}^2)^{1/2}}
, 
\nonumber \\
M_{ab}^{(N) 00} 
& = &
\left( 1 - \frac{T_a}{T_b} \right)
\frac{\alpha_{ab}}{(1+ \alpha_{ab}^2)^{3/2}}
, 
\nonumber \\
N_{ab}^{(0) 00} 
& = &  
- \frac{T_a}{T_b} M_{ab}^{(0)00}
= 
\frac{T_a}{T_b}
\frac{\alpha_{ab}}{(1+ \alpha_{ab}^2)^{1/2}}
, 
\nonumber \\
N_{ab}^{(N) 00} 
& = &
\alpha_{ab}^2 M_{ab}^{(N)00}
=
\left( 1 - \frac{T_a}{T_b} \right)
\frac{\alpha_{ab}^3}{(1+ \alpha_{ab}^2)^{3/2}}
.
\hspace{5mm}
\end{eqnarray}
Then, the $00$ elements, 
$M_{ab}^{00} = M_{ab}^{(0)00} + M_{ab}^{(N)00}$ and 
$N_{ab}^{00}  = N_{ab}^{(0)00} + N_{ab}^{(N)00}$,  which equal 
$M_{ab}^{(S)00}$ and $N_{ab}^{(S)00}$, respectively, 
are given by
\begin{eqnarray} 
\label{MNS00}
M_{ab}^{00} 
& =  & - N_{ab}^{00} 
= M_{ab}^{(S)00} 
= - N_{ab}^{(S)00} 
\nonumber \\ 
& =  &  
- \left( 1 + \frac{m_a}{m_b} \right)
\frac{\alpha_{ab}^3}{(1+ \alpha_{ab}^2)^{3/2}}
.
\end{eqnarray}

Next, the $0i$ elements  
$M_{ab}^{(A) 0i}$, $N_{ab}^{(A) 0i}$ $(A=0,N)$, 
$M_{ab}^{0i}$, and $N_{ab}^{0i}$ $(i=1,2)$  
are given by  
\begin{eqnarray} 
\label{Mab0i}
& & M_{ab}^{(0) 01} 
=
- \frac{3 \alpha_{ab}^3}{2 (1+ \alpha_{ab}^2)^{3/2}}
,
\nonumber \\
& & 
M_{ab}^{(0) 02} 
=
- \frac{15 \alpha_{ab}^5}{8 (1+ \alpha_{ab}^2)^{5/2}}
,
\nonumber \\
& & 
M_{ab}^{(N)01} 
=
(\theta_{ab}^2 - 1) M_{ab}^{(0)01}
= 
\left( 1 - \frac{T_a}{T_b} \right) 
\frac{3 \alpha_{ab}^3}{2 (1+ \alpha_{ab}^2)^{5/2}}
,
\nonumber \\
& & 
M_{ab}^{(N)02} 
=
(\theta_{ab}^2 - 1) M_{ab}^{(0)02}
= 
\left( 1 - \frac{T_a}{T_b} \right) 
\frac{15 \alpha_{ab}^5}{8 (1+ \alpha_{ab}^2)^{7/2}}
,
\nonumber \\
& & 
 M_{ab}^{01} 
=
\theta_{ab}^2 M_{ab}^{(0) 01} 
=
- \frac{3\alpha_{ab}^5}{2 (1+ \alpha_{ab}^2)^{5/2}}
\left( 1 + \frac{m_a}{m_b} \right)
,
\nonumber \\
& & 
 M_{ab}^{02} 
=
\theta_{ab}^2 M_{ab}^{(0) 02} 
=
- \frac{15 \alpha_{ab}^7}{8 (1+ \alpha_{ab}^2)^{7/2}}
\left( 1 + \frac{m_a}{m_b} \right)
,
\nonumber \\
& & 
 N_{ab}^{(0) 01} 
= 
- \frac{T_a}{T_b} \alpha_{ab} M_{ba}^{(0)01} 
=
\frac{T_a}{T_b} \frac{3 \alpha_{ab}}{2 (1+ \alpha_{ab}^2)^{3/2}}
,
\nonumber \\
& & N_{ab}^{(0) 02} 
= 
- \frac{T_a}{T_b} \alpha_{ab} M_{ba}^{(0)02} 
=
\frac{T_a}{T_b} \frac{15 \alpha_{ab}}{8 (1+ \alpha_{ab}^2)^{5/2}}
,
\nonumber \\
& & 
N_{ab}^{(N)01} 
=
- \frac{T_a}{T_b} \alpha_{ab} M_{ba}^{(N)01} 
=
\left( 1 - \frac{T_a}{T_b} \right) 
\frac{3 \alpha_{ab}^3}{2 (1+ \alpha_{ab}^2)^{5/2}}
,
\nonumber \\
& & 
N_{ab}^{(N)02} 
=
- \frac{T_a}{T_b} \alpha_{ab} M_{ba}^{(N)02} 
=
\left( 1 - \frac{T_a}{T_b} \right) 
\frac{15 \alpha_{ab}^3}{8 (1+ \alpha_{ab}^2)^{7/2}}
,
\nonumber \\
& & 
 N_{ab}^{01} 
=
 \frac{3\alpha_{ab}^3}{2 (1+ \alpha_{ab}^2)^{5/2}}
\left( 1 + \frac{m_a}{m_b} \right)
,
\nonumber \\
& & 
 N_{ab}^{02} 
=
 \frac{15\alpha_{ab}^3}{8 (1+ \alpha_{ab}^2)^{7/2}}
\left( 1 + \frac{m_a}{m_b} \right)
,
\end{eqnarray}
where $\theta_{ab}$ defined in Eq.~(\ref{theta_ab}) and 
the momentum conservation law shown in Eq.~(\ref{mom_A}) 
are used. 

The $i0$ elements $M_{ab}^{(A) i0}$ and $N_{ab}^{(A) i0}$ 
$(A=0,N; i=1,2)$ are given by  
\begin{eqnarray} 
\label{MA10}
& & M_{ab}^{(0) 10} 
=
M_{ab}^{(0) 01}
,
\hspace{5mm}
 M_{ab}^{(0) 20} 
=
M_{ab}^{(0) 02}
,
\nonumber \\
& & 
M_{ab}^{(N)10} 
=
\left( \frac{T_a}{T_b}  - 1 \right) 
\frac{\alpha_{ab} ( 10 + \alpha_{ab}^2 )}{2 (1+ \alpha_{ab}^2)^{5/2}}
,
\nonumber \\
& & 
M_{ab}^{(N)20} 
=
\left( \frac{T_a}{T_b}  - 1 \right) 
\frac{3\alpha_{ab}^3 ( 28 + 3 \alpha_{ab}^2 )}{8 (1+ \alpha_{ab}^2)^{7/2}}
,
\nonumber \\
& & 
N_{ab}^{(0) 10} 
= 
\frac{T_a^2}{T_b^2} \alpha_{ab} N_{ba}^{(0)01} 
=
- \frac{T_a}{T_b} M_{ab}^{(0)01} 
,
\nonumber \\
& & 
N_{ab}^{(0) 20} 
= 
\frac{T_a^2}{T_b^2} \alpha_{ab} N_{ba}^{(0)02} 
=
- \frac{T_a}{T_b} M_{ab}^{(0)02} 
,
\nonumber \\
& & 
N_{ab}^{(N)10} 
=
\left( 1 - \frac{T_a}{T_b} \right) 
\frac{3 \alpha_{ab}^3 ( - 2 + \alpha_{ab}^2 )}{2 (1+ \alpha_{ab}^2)^{5/2}}
,
\nonumber \\
& & 
N_{ab}^{(N)20} 
=
\left( 1 - \frac{T_a}{T_b} \right) 
\frac{15 \alpha_{ab}^5 ( - 4 + \alpha_{ab}^2 )}{8 (1+ \alpha_{ab}^2)^{7/2}}
,
\end{eqnarray}
where the relations shown in Eqs.~(\ref{mom_A}) and (\ref{symmetry0}) 
are used. 
Using Eqs.~(\ref{Mab0i}) and (\ref{MA10}), 
we can immediately evaluate 
$M_{ab}^{i0} = M_{ab}^{(0)i0} + M_{ab}^{(N)i0}$ and 
$N_{ab}^{i0} = N_{ab}^{(0)i0} + N_{ab}^{(N)i0}$ $(i=1,2)$. 

The $ij$ elements $M_{ab}^{(A) ij}$ and $N_{ab}^{(A) ij}$ 
$(A=0,N; i=1,2)$ are written as  
\begin{eqnarray} 
\label{Mabij}
& & M_{ab}^{(0) 11} 
=
-
\frac{\alpha_{ab}( 30 + 16 \alpha_{ab}^2 + 13 \alpha_{ab}^4 )}{
4 ( 1 + \alpha_{ab}^2 )^{5/2}}
,
\nonumber \\
& & 
 M_{ab}^{(0)12} = M_{ab}^{(0)21} 
=
- \frac{3\alpha_{ab}^3 ( 84 + 32 \alpha_{ab}^2 + 23 \alpha_{ab}^4 )}{
16 (1+ \alpha_{ab}^2)^{7/2}}
,
\nonumber \\
& & 
M_{ab}^{(0)22} 
=
- 
 \frac{\alpha_{ab} }{
64(1+ \alpha_{ab}^2)^{9/2}}
\nonumber \\
& & \mbox{} 
\hspace*{3mm}
\times
( 1400 + 1792 \alpha_{ab}^2 + 3672 \alpha_{ab}^4 
+ 1088 \alpha_{ab}^6 + 433 \alpha_{ab}^8)
,
\nonumber \\
& & 
M_{ab}^{(N)11} 
=
\left(  1 - \frac{T_a}{T_b} \right) 
\frac{3 \alpha_{ab} ( 10 - 2 \alpha_{ab}^2  + 3 \alpha_{ab}^4 )}{4 (1+ \alpha_{ab}^2)^{7/2}}
,
\nonumber \\
& & 
 M_{ab}^{(N)12} 
=
\left(  1 - \frac{T_a}{T_b} \right) 
 \frac{3\alpha_{ab}^3 ( 84 - 2 \alpha_{ab}^2 + 19 \alpha_{ab}^4 )}{
16 (1+ \alpha_{ab}^2)^{9/2}}
,
\nonumber \\
& & 
 M_{ab}^{(N)21} 
=
- \left(  1 - \frac{T_a}{T_b} \right) 
 \frac{\alpha_{ab}}{
16 (1+ \alpha_{ab}^2)^{9/2}}
\nonumber \\
& & 
\mbox{} 
\hspace*{16mm} \times
 ( 280 + 84 \alpha_{ab}^2 + 348 \alpha_{ab}^4
+ 19\alpha_{ab}^6 )
, 
\nonumber \\
& &  M_{ab}^{(N)22} 
= 
\left(  1 - \frac{T_a}{T_b} \right) 
 \frac{\alpha_{ab}}{
64 (1+ \alpha_{ab}^2)^{11/2}}
\nonumber \\
& & 
\hspace*{3mm}
 \times
 ( 1400 - 112 \alpha_{ab}^2 + 2424 \alpha_{ab}^4
-556 \alpha_{ab}^6 + 233 \alpha_{ab}^8 )
, 
\nonumber \\
& & 
N_{ab}^{(0) 11} 
= 
\frac{T_a}{T_b} 
\frac{27 \alpha_{ab}^3}{4 (1+ \alpha_{ab}^2)^{5/2}}
,
\nonumber \\
& & N_{ab}^{(0) 12} 
= 
\frac{T_a}{T_b} 
\frac{225 \alpha_{ab}^3}{16 (1+ \alpha_{ab}^2)^{7/2}}
,
\nonumber \\
& & N_{ab}^{(0) 21} 
= 
\frac{T_a}{T_b} 
\frac{225 \alpha_{ab}^5}{16 (1+ \alpha_{ab}^2)^{7/2}}
,
\nonumber \\
& & N_{ab}^{(0) 22} 
= 
\frac{T_a}{T_b} 
\frac{2125 \alpha_{ab}^5}{64 (1+ \alpha_{ab}^2)^{9/2}}
,
\nonumber \\
& & 
N_{ab}^{(N)11} 
=
\left( 1 - \frac{T_a}{T_b} \right) 
\frac{9 \alpha_{ab}^3 ( - 2 + 3 \alpha_{ab}^2 )}{4 (1+ \alpha_{ab}^2)^{7/2}}
,
\nonumber \\
& & 
N_{ab}^{(N)12} 
=
\left( 1 - \frac{T_a}{T_b} \right) 
\frac{45 \alpha_{ab}^3 ( - 2 + 5 \alpha_{ab}^2 )}{16 (1+ \alpha_{ab}^2)^{9/2}}
, 
\nonumber \\
& & 
N_{ab}^{(N)21} 
=
\left( 1 - \frac{T_a}{T_b} \right) 
\frac{75 \alpha_{ab}^5 ( - 4 + 3 \alpha_{ab}^2 )}{16 (1+ \alpha_{ab}^2)^{9/2}}
, 
\nonumber \\
& & 
N_{ab}^{(N)22} 
=
\left( 1 - \frac{T_a}{T_b} \right) 
\frac{525 \alpha_{ab}^5 ( - 4 + 5 \alpha_{ab}^2 )}{64 (1+ \alpha_{ab}^2)^{11/2}}
.
\hspace*{8mm}
\end{eqnarray}
Then,  $M_{ab}^{ij} = M_{ab}^{(0)ij} + M_{ab}^{(N)ij}$ and 
$N_{ab}^{ij} = N_{ab}^{(0)ij} + N_{ab}^{(N)ij}$ $(i,j=1,2)$ 
are evaluated from the results shown in Eq.~(\ref{Mabij}). 

Now, we can use the matrix elements
$M_{ab}^{(0)0i} = M_{ab}^{(0)i0}$ and $M_{ab}^{(0)ij}$ 
$(i,j=1,2, \cdots)$  
to express
the matrix elements
$M_{ab}^{(S)0i} = M_{ab}^{(S)i0}$ and $M_{ab}^{(S)ij}$ 
$(i,j=1,2, \cdots)$  
by
\begin{eqnarray} 
\label{MS0i}
& & 
M_{ab}^{(S)0i} 
=
M_{ab}^{(S)i0} 
=
\theta_{ab}  M_{ab}^{(0)0i}
,
\nonumber \\
& & 
M_{ab}^{(S)ij} 
=
M_{ab}^{(0)ij}
, 
\end{eqnarray}
and write
the matrix elements 
$N_{ab}^{(S)0i}$ and $N_{ab}^{(S)i0}$ $(i=1,2, \cdots)$ 
as 
\begin{eqnarray} 
& & 
N_{ab}^{(S)0i} 
= 
- \frac{T_a}{T_b} \alpha_{ab} M_{ba}^{(S)0i}
=
\theta_{ba}
N_{ab}^{(0)0i}
,
\nonumber \\
& & 
N_{ab}^{(S)i0} 
= 
\frac{T_a}{T_b} \alpha_{ab} N_{ba}^{(S)0i}
=
-
M_{ab}^{(S)0i}
, 
\end{eqnarray}
where Eqs.~(\ref{MS0j}), (\ref{MNLS}), (\ref{mom_A}), and 
(\ref{MS0i}) are used. 
Then, Eq.~(\ref{NabSij}) can be used to
evaluate $N_{ab}^{(S)ij}$ $(i,j=1,2,\cdots)$ from 
 $N_{ab}^{(S)i0}$, $N_{ab}^{(S)0j}$ and 
$N_{ab}^{00}$ [see Eq.~(\ref{MNS00})].

\section{IMPROVED SUGAMA OPERATOR MODIFIED BY SYMMETRIZING 
MATRIX ELEMENTS}

In this Appendix, the improved Sugama operator defined 
in Eq.~(\ref{imp}) is modified when $T_a \neq T_b$ as 
follows:  
\begin{equation}
\label{imp*}
C_{ab}^{LS* {\rm (imp)}} (\delta f_a, \delta f_b)
\equiv 
C_{ab}^{LS} (\delta f_a, \delta f_b)
+  \Delta C_{ab}^{F*} (\delta f_b)
\end{equation}
where $C_{ab}^{LS} (\delta f_a, \delta f_b)$ represents the 
original Sugama operator described in Sec.~III and 
the new correction part $\Delta C_{ab}^{F*}(\delta f_b)$ is defined by 
\begin{equation} 
\label{DCab*}
 \Delta C_{ab}^{F*} (\delta f_b)
 \equiv
f_{aM} \frac{m_a}{T_a} {\bf v} \cdot  
\sum_{j=1}^{\infty} \Delta {\bf C}_{ab j}^{F*} [\delta f_b]
L_j^{(3/2)}(x_a^2)
.
\end{equation}
Here, $\Delta {\bf C}_{ab j}^{F*}[\delta f_b]$ 
$(j = 1, 2, \cdots)$ are given by  
\begin{equation}
\label{DCabj*}
\Delta {\bf C}_{ab j}^{F*} [\delta f_b] 
\equiv  
\frac{c_j}{\tau_{ab}} 
\sum_{k=1}^{\infty} 
\Delta N_{ab}^{*jk} \; {\bf u}_{b k}[\delta f_b]
\hspace*{5mm}
(j=1, 2, \cdots)
, 
\end{equation}
and 
\begin{eqnarray}
& & 
\Delta N_{ab}^{*jk} 
 \equiv  
N_{ab}^{*jk} - N_{ab}^{(S) jk}
\nonumber \\ 
& & 
=
\frac{N_{ab}^{00} N_{ab}^{*jk} - N_{ab}^{(S)i0} N_{ab}^{(S)0j} }{N_{ab}^{00}} 
\hspace*{2mm}
(j, k =1, 2, \cdots)
,  
\hspace*{2mm}
\end{eqnarray}
where 
\begin{equation}
\label{Nab*}
N_{ab}^{*jk}
\equiv
\frac{ T_a v_{Ta}}{2}
\left( 
\frac{N_{ab}^{jk}}{T_a v_{Ta}}+ 
\frac{N_{ba}^{kj}}{T_b v_{Tb}}
\right)
\hspace*{2mm}
(j, k =1, 2, \cdots)
. 
\end{equation}

We can now use the test and field particle part of 
$C_{ab}^{LS* {\rm (imp)}} (\delta f_a, \delta f_b)$ to 
obtain the matrix elements $M_{ab}^{*ij}$ and 
$N_{ab}^{*ij}$ in the same way as shown in Eq.~(\ref{MN}). 
Then, the friction coefficients $l_{*ij}^{ab}$ can be  
derived from $M_{ab}^{*ij}$ and $N_{ab}^{*ij}$ 
[see Eq.~(\ref{lij})].  
Since $\Delta C_{ab}^{F*}(\delta f_b)$ 
defined in Eq.~(\ref{DCab*}) gives the correction only in the 
field particle part, we immediately see that  
\begin{equation}
\label{Mab*}
M_{ab}^{*ij} = M_{ab}^{(S) ij}
\hspace*{2mm}
(i, j =0, 1, 2, \cdots)
. 
\end{equation}
We also find that 
\begin{equation}
\label{Nab*0ij}
N_{ab}^{*i0} = N_{ab}^{(S)i0}, 
\hspace*{2mm}
N_{ab}^{*0j} = N_{ab}^{(S)0j}, 
\hspace*{2mm}
(i, j =0, 1, 2, \cdots)
, 
\end{equation}
and 
$N_{ab}^{*ij}$ $(i, j = 1, 2, \cdots)$ are given by Eq.~(\ref{Nab*}). 
It is confirmed from Eqs.~(\ref{Nab*}), (\ref{Mab*}) and (\ref{Nab*0ij}) 
that the matrix elements $M_{ab}^{*ij}$ and $N_{ab}^{*ij}$ 
 satisfy 
\begin{equation}
\label{MN*_sym}
M_{ab}^{*ij}  =  M_{ab}^{*ji}, 
\; \; 
\frac{N_{ab}^{*ij}}{T_a v_{Ta}}  =  
\frac{N_{ba}^{*ji}}{T_b v_{Tb}} 
\; \; 
(i, j = 0, 1, 2, \cdots)
,
\end{equation}
which leads to the symmetry of the friction coefficients $l_{*ij}^{ab}$,
\begin{equation}
\label{l*sym}
l_{*ij}^{ab}=l_{*ji}^{ba}
\; \; 
(i, j = 1, 2, \cdots)
. 
\end{equation}
The modified operator 
$C_{ab}^{LS* {\rm (imp)}} (\delta f_a, \delta f_b)$
exactly satisfies the adjointness relations 
in the same form as those in Eq.~(\ref{adjoint}) 
and accordingly induces the Onsager symmetry of 
collisional transport coefficients. 

When $T_a \neq T_b$,  
the values of $M_{ab}^{*ij}$, $N_{ab}^{*ij}$, and $l_{*ij}^{ab}$ 
are different from those of $M_{ab}^{ij}$, $N_{ab}^{ij}$, and $l_{ij}^{ab}$ 
given by the linearized Landau operator, respectively.  
However, as explained in Appendix~B, the differences between these values 
are not expected to cause serious errors in solutions of kinetic equations  
because $m_a/m_b \ll 1$ or $m_a/m_b \gg 1$ 
are required if $T_a$ and $T_b$ differ significantly from each other. 

Noting that $\Delta C_{ab}^{F*}(\delta f_b)$ 
never influences collisional momentum and energy transfer, 
we can confirm that 
$C_{ab}^{LS* {\rm (imp)}}(\delta f_a, \delta f_b)$
keeps conservation laws of particles, momentum, and energy. 
Especially, the momentum conservation law imposes 
the constraints on the matrix elements and the friction coefficients as  
\begin{eqnarray}
& & 
M_{ab}^{*0j} + 
\frac{T_a v_{Ta}}{T_b v_{Tb}} N_{ba}^{*0j}
= 0
\hspace*{3mm} ( j = 0, 1, 2, \cdots),
\nonumber \\ & & 
\sum_a l_{*1j}^{ab} = 0
\hspace*{3mm} 
( j = 1, 2, \cdots),  
\end{eqnarray}
which are rewritten with the help of Eqs.~(\ref{MN*_sym}) and (\ref{l*sym}) 
as  
\begin{eqnarray}
\label{MN*j0}
& & 
M_{ab}^{*j0} + 
N_{ab}^{*j0}
= 0
\hspace*{3mm} ( j = 0, 1, 2, \cdots),
\nonumber \\ & & 
\sum_b l_{*j1}^{ab} = 0
\hspace*{3mm} 
( j = 1, 2, \cdots).   
\end{eqnarray}
Then, if the perturbed functions are written as
$
\delta f_s
=
f_{sM} (m_a/T_s) ( {\bf u}_s \cdot {\bf v} )
$
$(s = a, b)$
with the condition ${\bf u}_a = {\bf u}_b$, 
we find that 
$C_{ab}^{LS* {\rm (imp)}} (\delta f_a, \delta f_b)
= \tau_{ab}^{-1} f_{aM} (m_a/T_a) {\bf v} \cdot 
\sum_{j=0}^\infty c_j L_j^{(3/2)} (x_a^2)
(M_{ab}^{*j0} {\bf u}_a + N_{ab}^{*j0} {\bf u}_b) 
= 0$ 
because of Eq.~(\ref{MN*j0}) and ${\bf u}_a = {\bf u}_b$. 
Noting that $C_{ab}^{LS* {\rm (imp)}}$ is also annihilated by 
the perturbed distribution functions 
$\delta f_s$ $(s=a,b)$ given by Eq.~(\ref{shiftM0}) with 
$\delta T_a / T_a = \delta T_b/T_b$,  
it is now remarked that 
$C_{ab}^{LS* {\rm (imp)}} (\delta f_a, \delta f_b)$ vanishes 
for the perturbed distribution functions 
given by 
the perturbed Maxwellian with 
the perturbed densities $\delta n_s$, 
temperatures $\delta T_s$, and 
flows ${\bf u}_s$ $(s=a,b)$ as 
\begin{equation}
\delta f_s
=
f_{sM}
\left[ 
\frac{\delta n_s}{n_s}
+ \frac{m_s}{T_s} {\bf u}_s \cdot {\bf v}
+ \frac{\delta T_s}{T_s}
\left(
\frac{m_s v^2}{2T_s} - \frac{3}{2}
\right)
\right]  
,
\end{equation}
where 
${\bf u}_a = {\bf u}_b$
and 
$\delta T_a / T_a = \delta T_b / T_b$. 

Using Eq.~(\ref{imp*}), 
the collision operator for gyrokinetic equations is given by 
\begin{eqnarray}
\label{CGK*}
& & 
\oint \frac{d \xi}{2\pi}
e^{i {\bf k}_\perp \cdot \mbox{\boldmath $\rho$}_a}
C_{ab}^{LS*{\rm (imp)}}(
e^{-i {\bf k}_\perp \cdot
\mbox{\boldmath $\rho$}_a} h_{a {\bf k}_\perp}, 
e^{-i {\bf k}_\perp \cdot
\mbox{\boldmath $\rho$}_b} h_{b {\bf k}_\perp}
)
\nonumber \\ 
& = & 
\oint \frac{d \xi}{2\pi}
e^{i {\bf k}_\perp \cdot \mbox{\boldmath $\rho$}_a}
C_{ab}^{LS}(
e^{-i {\bf k}_\perp \cdot
\mbox{\boldmath $\rho$}_a} h_{a {\bf k}_\perp}, 
e^{-i {\bf k}_\perp \cdot
\mbox{\boldmath $\rho$}_b} h_{b {\bf k}_\perp}
)
\nonumber \\ 
&  & 
\mbox{} + 
\oint \frac{d \xi}{2\pi}
e^{i {\bf k}_\perp \cdot \mbox{\boldmath $\rho$}_a}
\Delta C_{ab}^{F*}(
e^{-i {\bf k}_\perp \cdot
\mbox{\boldmath $\rho$}_a} h_{a {\bf k}_\perp}, 
e^{-i {\bf k}_\perp \cdot
\mbox{\boldmath $\rho$}_b} h_{b {\bf k}_\perp}
)
.
\nonumber \\ & & 
\end{eqnarray}
The detailed expression of the first term on the right-hand of 
Eq.~(\ref{CGK*}) is found in Ref.~\cite{Sugama2009} 
while the second term is expressed by Eq.~(\ref{DCGK0}) with 
putting $\Delta M_{ab}^{jk}=0$ and replacing 
$\Delta N_{ab}^{jk}$ by 
$\Delta N_{ab}^{*jk} \equiv N_{ab}^{*jk} - N_{ab}^{(S)jk}$. 

Since the two colliding particle species need to have very different masses for their temperatures to be significantly different from each other, 
the improved Sugama operators presented in this Appendix and Sec.~IV do not seem to 
show large quantitative differences from each other for the case of $T_a \neq T_b$ 
where $m_a/m_b \ll 1$ or $m_a/m_b \gg 1$ holds. 
It is not so clear how the adjointness properties of the linearized collision operator 
is crucial for accurate prediction of turbulent transport 
or for formulation of efficient turbulence simulation methods 
in comparison with their roles in neoclassical transport theory and simulation. 
Unless one can recognize merits of the adjointness properties 
for analytical or numerical calculations of turbulent transport, 
the operator presented in Sec.~IV may seem more suitable for gyrokinetic simulation in the unequal temperature case than that 
in this Appendix because the former describes the friction-flow relations 
more accurately. 
However, we still note that there are several theoretical studies on the Onsager-type symmetry of the quasilinear turbulent transport 
matrix,~\cite{Horton1980,Sugama1995,Sugama1996b,Garbet2012,SWang} 
for which the collision operator 
given in this Appendix can be useful to study collisional effects. 



\end{document}